\documentclass[12pt]{article}
\usepackage{rotate}
\usepackage{epsfig}
\usepackage{psfrag}
\usepackage{a4}

\begin{document}

\newcommand \be  {\begin{equation}}
\newcommand \bea {\begin{eqnarray} \nonumber }
\newcommand \ee  {\end{equation}}
\newcommand \eea {\end{eqnarray}}

\title{\bf Relation between Bid-Ask Spread, Impact and Volatility in Order-Driven Markets}

\author{Matthieu Wyart$^{*,+}$, Jean-Philippe Bouchaud$^{*}$, Julien Kockelkoren$^{*}$, \\ 
Marc Potters$^*$, Michele Vettorazzo$^*$}
\maketitle
{\small
{$*$ Science \& Finance, Capital Fund Management, 6 Bvd Haussmann},
{75009 Paris, France}\\
{$+$ now at: Division of Engineering and Applied Sciences, Harvard University,
Pierce Hall, 29 Oxford Street, Cambridge, Massachusetts 02138, USA}
}

\begin{abstract}
We show that the cost of market orders and the profit of infinitesimal market-making or -taking strategies can be expressed 
in terms of directly observable quantities, namely the spread and the lag-dependent impact function. Imposing that any market taking 
or liquidity providing strategies is at best marginally profitable, we obtain a linear relation between 
the bid-ask spread and the instantaneous impact of market orders, in good agreement with our empirical observations on 
{\it electronic} markets. We then use this 
relation to justify a strong, and hitherto unnoticed, empirical correlation between the spread and the volatility {\it per trade}, 
with $R^2$s exceeding $0.9$. 
This correlation suggests both that the main determinant of the bid-ask spread is adverse selection, and that most of the volatility
comes from trade impact. We argue that the role of
the time-horizon appearing in the definition of costs is crucial and that long-range correlations in the order flow, overlooked in 
previous studies, must be carefully factored in. We find that the spread is significantly larger on the {\sc nyse}, a liquid market with 
specialists, where monopoly rents appear to be present.
\end{abstract}

\section{Introduction and review of the literature}
\label{intro}

One of the most important attribute of financial markets is to provide immediate liquidity to investors \cite{Orlean}, 
who are able to convert cash into stocks and vice-versa nearly instantaneously whenever they choose to do so. Of course, 
some markets are more liquid than others and the liquidity of a given market varies in time and can in fact
dramatically dry up in crisis situations. How should markets be organized, at the micro-structural level, to 
optimize liquidity, to favor steady and orderly trading and avoid these liquidity crises? In the past, 
the burden of providing liquidity was given to ``market makers'' (or specialists). In order to ensure steady trading, 
the specialists alternatively sell to buyers and buy to sellers, and get compensated by the so-called 
bid-ask spread -- i.e. the price at which they sell to the crowd is always slightly larger than the price at which 
they buy. The determinants of the value of the spread in specialists markets have been the subject of many studies 
in the economics literature \cite{OH,BFH,Mad,Glo1,Glo2,Has,HS,MadS,Holli}, and \cite{Stoll} for a recent review. 

However, most financial markets have nowadays become fully electronic 
(with the notable exception of the New-York Stock Exchange, {\sc nyse} -- although this will soon change). 
In these markets, liquidity is {\it self-organized}, 
in the sense that any agent can choose, at any instant of time, to either provide liquidity or consume liquidity. 
More precisely, any agent can provide liquidity by posting {\it limit orders}: these are propositions to sell 
(or buy) a certain volume of shares or lots at a fixed minimum (maximum) price. Limit orders are stored in 
the order-book. At a given instant in time, the best offer on the sell side (the `ask') is higher than the 
best price on the buy side (the `bid') so no transaction takes place.  For a transaction to occur, an agent must 
consume liquidity by issuing a {\it market order} to buy (or to sell) a certain number of shares; the transaction occurs 
at the best available price, provided the volume in the order book at that price is enough to absorb the incoming market 
order. Otherwise, the price `walks up' (or down) the ladder of offers in the order book, until the order is fully 
satisfied. The liquidity of the market is partially characterized by the bid-ask spread $S$, which sets the cost of an 
instantaneous round-trip of one share (a buy instantaneously followed by a sell, or vice versa).\footnote{Other determinants 
of liquidity discussed in the literature are the depth of the order book and market resiliency, see \cite{Black, Kyle}.} 
A liquid market is such that this cost is small. A question of both theoretical and practical crucial importance 
is to know what fixes the magnitude of the spread in the self-organized set-up of electronic markets, and the relative
merit of limit vs. market orders. In the present work, we argue that on electronic markets, profitable high frequency 
strategies using either limit or market orders should not exist, imposing a linear relation between the bid-ask spread $S$ 
and the average impact of market orders. This, in turn justifies a simple, but hitherto unnoticed, proportionality relation 
between the spread and the volatility {\it per trade}.

In a large fraction of the economics literature \cite{OH,BFH,Mad,Glo1}, liquidity providers are 
described as market makers who earn their profit from the spread. The value of the spread is non zero because 
this market making strategy has costs. Three types of cost are discussed in the literature \cite{Stoll}: 
\begin{itemize}
\item (i) order processing costs (which includes sheer profit for the market maker); 
\item (ii) adverse selection costs: liquidity takers may have superior information on the future price of the stock, in which case the 
market maker loses money;\footnote{This is also discussed as the free option trading problem in the literature, see 
e.g. \cite{Liu} and refs. therein.} 
\item (iii) inventory risk: market makers may temporarily accumulate large long or short positions which
are risky. If agents are risk-sensitive and have to limit their exposure, this adds extra-costs.
\end{itemize}
Theoretical models that account for these costs typically 
introduce a rather large amount of free parameters (such as risk-aversion, fraction of informed trades, fraction of patient/
impatient traders, etc.) most of which cannot be measured directly. In order to extract the different determinants of the 
spread from empirical data, some drastic assumptions must be made. For example, assuming the order flow to 
be short-ranged correlated, Huang and Stoll \cite{HS} find (using data from 1992) that $90 \%$ of the spread is associated to 
order processing 
costs, and not to adverse selection\footnote{Adverse selection is even found to have, within this framework, a negative contribution 
to the spread!}. This is would mean rather comfortable profits for market making\footnote{Direct processing costs can be estimated to 
be at least ten times smaller than the spread, in particular on electronic markets.}, and is a somewhat surprising conclusion 
since the spread on purely electronic markets is found comparable to the spread in markets with specialists. A related approach 
is that of \cite{MadS}, where the ratio of adverse selection to processing costs was estimated to 
be in the range $35-50 \%$ on the {\sc nyse} in 1990 (see also \cite{Stoll} for similar numbers). We will review this theoretical 
framework below and detail the similarities and 
differences with our own analysis; one particularly crucial difference is the assumption that the order imbalance has 
short-ranged correlations \cite{HS,MadS}, 
and therefore that market impact of a single trade is {\it permanent}, in striking disagreement with empirical data, 
where the order flow is instead found to be a long-memory process \cite{QF04,Farmer1}, and single trade impact {\it transient}, 
but decaying 
very slowly \cite{QF04,QF05}. The long-range correlation between trades, and the corresponding temporal dependence of market impact 
will turn out to play an important role in the following discussion.

On general grounds, both adverse selection and inventory risk imply a positive correlation between the spread and the volatility of 
the traded asset. This makes perfect intuitive sense, and the aim of the present paper is to clarify in detail the origin of this
relation. Positive correlation between spread and volatility is indeed documented empirically 
(see e.g. \cite{Bess,Mad,Cop,Stoll,Chordia1,Chordia2,Chordia3,Farmer0}), 
but is not particularly spectacular and stands as one among other reported correlations, e.g., with traded volume, flow of 
limit orders, market capitalization, etc.\cite{Stoll}. Here, we want to argue theoretically, and demonstrate empirically on 
different markets, that there is in fact a very strong correlation between the spread and the volatility {\it per trade}, 
rather than with the volatility {\it per unit time}. Such a strong relation was first noted on the case of France-Telecom 
\cite{QF04}, and independently on the stocks of the {\sc ftse}-100 \cite{Zumbach}, but no theoretical argument was given 
in favor of this relation. 

From a theoretical point of view, several statistical models of limit and market order flows have been analyzed to
understand the distribution of the bid-ask spread, and relate its average value to flow and cancellation rates 
\cite{BFH,Fou1,Fou2,Farmer2,BMP,Farmer3,Luck,Io}. Some 
models include strategic considerations in order placement and look for a trade-off between the cost of delayed 
execution and that of immediacy, but suppose that the price dynamics is bounded in a finite interval \cite{Fou2}, therefore 
neglecting the long term volatility of the price (see also \cite{Luck,Io}). As such, these finite band models have 
nothing to say about 
the spread-volatility relationship. Another line of models discards all strategic components (``Zero intelligence models'') 
and assume Poisson rates for limit orders, market orders and cancellation \cite{Farmer2,BMP,Farmer3}.\footnote{More
elaborated `weak intelligence' models have been studied recently, see \cite{Mike}.}
One can then compute both the average bid-ask spread and the long-term volatility as a function of these Poisson rates, 
and compare these predictions with empirical data \cite{Farmer4}. The
problem with such models is that although the order flow itself is completely random, the persistence of the order book 
leads to strong non-diffusive short term predictability of the {\it price}, which would be very easily picked off by high frequency 
automated execution machines. These programs search to optimize execution costs (see e.g. \cite{Holli,Almgren}) by adequately 
conditioning the order flow 
(proportion of limit and market orders, timing, aggressivity) and use any short-term predictability to do so. 
As a result there are in fact very strong high frequency correlations in the order flow, coming from the `hide and seek' game played 
by buyers and sellers within the order book \cite{QF04,Farmer1,Weber}. A key observation is that for small tick stocks, 
the total available volume in the order book at any instant in time is in fact {\it extremely small}, on the order of 
$10^{-5}-10^{-4}$ of the market capitalization, or $10^{-3}-10^{-2}$ of the daily volume (see Table 2 in Appendix 2). 
Clearly, the reason for such a small outstanding liquidity is that liquidity providers want to avoid giving a free trading
option to informed traders. As a consequence, liquidity takers must cut their total order in small
chunks; this creates the long term correlation in  order flow \cite{FarmerL}. But since on electronic markets 
sophisticated buyers and sellers can trade using at their best convenience either limit or market orders, 
the average cost of limit and market orders should be very similar. If -- say -- market orders were on average significantly 
more expensive than limit orders, more limit orders would be issued, thereby reducing the spread and the cost of market orders, 
until an equilibrium is reached.\footnote{Data from brokers VWAP machines indeed show that the fraction of issued market orders is close
to $50 \%$.} That a competitive ecology between limit and market orders should exist on order-driven markets was 
emphasized in \cite{Handa,QF04,QF05}. However, as our analysis reveals, this ecology turns out to be 
considerably more intricate than anticipated by Handa \& Schwartz \cite{Handa}.

In the following, we introduce the idea of infinitesimal strategies, participating to a vanishing fraction of market or
limit orders. Imposing that such strategies lead at best to marginal profits motivates a linear relation between the instantaneous 
price impact of a market order and the bid-ask spread, which we check empirically. Interestingly, we find that the profitability
of these strategies depend in a non trivial way on the time horizon over which they are implemented. We show in particular 
that fast market making strategies can be profitable even though the long-term average cost of limit orders is positive, a rather
paradoxical situation brought about by the presence of long-range correlations in the order flow and the temporal structure
of the impact function.

The linear relation between spread and impact in turn allows us to establish a proportionality 
relation between the spread and volatility {\it per trade}, which holds both across different stocks and for a given stock
across time, on electronic markets and on the {\sc nyse}. This result shows that in a competitive electronic market the 
bid-ask spread in fact mostly comes from ``adverse selection'', provided one extends this notion to account for the 
fact that trades can be uninformed but still impact the price. What is relevant here is that {\it any} unexpected component 
of the market order flow, whether it is truly informed or just random, impacts the price and creates a cost for limit 
orders, which must be compensated by the spread, as we now explain in detail.  

\section{Limit orders {\it vs} market orders and market impact}
\label{s2}

\subsection{A simple theoretical framework}

We start by reviewing the theoretical framework proposed by Madhavan, Richardson and Roomans ({\sc mrr}) in \cite{MadS}, which helps 
define various quantities and hone in on relevant questions. We will call $v_i$ the volume of the $i$th market order, 
and $\epsilon_i$ the sign of that market order ($\epsilon=+1$ for a buy and $\epsilon=-1$ for a sell). The assumptions of the model 
are (i) that all trades have the same volume $v_i=v$ and (ii) the $\epsilon_i$'s are generated by a Markov process with 
correlation $\rho$, which means that the average value of $\epsilon_i$ conditioned on the past only depends on $\epsilon_{i-1}$ and is
given by:
\be
\left.\langle \epsilon_i \rangle\right|_{\epsilon_{i-1}} = \rho \epsilon_{i-1},
\ee
where $\langle ... \rangle$ denotes averaging. 
The case $\rho=0$ corresponds to independent trade signs, whereas $\rho > 0$ describes positive autocorrelations of trades. Note
that in this model, correlations decay exponentially:
\be
C(\ell) = \langle \epsilon_i \epsilon_{i+\ell} \rangle = \rho^\ell.
\ee
The {\sc mrr} model assumes that the `true' price $p_i$ evolves both because of random external shocks (or news) and because of trade impact. 
It is natural to postulate that both external news and surprise in order flow should move the price. Since the surprise at the $i$th 
trade is given by $\epsilon_i -  \rho \epsilon_{i-1}$, {\sc mrr} write the following 
evolution equation for the price:
\be
p_{i+1} - p_i = \xi_i + \theta [\epsilon_i -  \rho \epsilon_{i-1}],
\ee
where $\xi$ is the shock component, with variance $\langle \xi_i^2 \rangle= \Sigma^2$, and $\theta$ measures trade impact, assumed to
be constant (all trades are assumed to have the same volume). Since market makers cannot guess the surprise of the next trade, 
they post a bid price $b_i$ and an ask price $a_i$ given by:
\be
a_i = p_i + \theta [1 -  \rho \epsilon_{i-1}] + \phi; \qquad b_i = p_i + \theta [-1 -  \rho \epsilon_{i-1}] - \phi,
\ee
where $\phi$ is the extra compensation claimed the market maker, covering processing costs and the shock component risk. 
The above rule ensures no {\it ex-post} regrets for the market maker. The 
spread is therefore $S \equiv a -b =2(\theta + \phi)$, whereas the midpoint $m \equiv (a+b)/2$ immediately before the $i$th trade 
is given by:
\be
m_i = p_i - \theta \rho \epsilon_{i-1}.
\ee
These equations allow to compute several important quantities for the following discussion, although not explicitely considered 
by {\sc mrr}. The first one is the lagged impact 
function introduced in \cite{QF04,QF05}:
\be
{\cal R}_\ell = \langle \epsilon_i \cdot (m_{\ell+i}-m_i)\rangle,
\ee 
which is found, within the {\sc mrr} model, to increase from ${\cal R}_1=\theta(1-\rho)$ to ${\cal R}_\infty=\theta$ (See Appendix 1
and Fig. 1).
Due to correlations between trades, the long time impact is therefore enhanced compared to the short term 
impact by a factor:
\be
\label{enhance}
\lambda_\infty=\frac{1}{1-C_1},
\ee
where $C_1=C(\ell=1)=\rho$ in the {\sc mrr} model, but the above relation is more general (see Appendix I). 

The second quantity is the mid-point volatility, defined as:
\be
\sigma^2_\ell = \frac{1}{\ell} \langle (m_{\ell+i}-m_i)^2 \rangle,
\ee
which is easily computed to be:\footnote{The is an extra contribution to $\sigma_1^2$ coming from any high-frequency noise 
component that we neglect here, coming from decimalisation, small volumes at bid/ask, etc. See \cite{MadS,QF04} and footnote 18
below.}
\be\label{sigmamrr}
\sigma_1^2 = {\cal R}_1^2 + \Sigma^2; \qquad  \sigma_\infty^2 = \frac{1+\rho}{1-\rho} {\cal R}_1^2 + \Sigma^2; \qquad
\langle \xi_i^2 \rangle= \Sigma^2.
\ee
Within the above interpretation, the {\sc mrr} model leads to the following simple relations between spread, 
impact and volatility per trade: 
\be\label{spreadmrr}
S = 2 \lambda_\infty {\cal R}_1 + 2 \phi \qquad \sigma_1^2 = {\cal R}_1^2 + \Sigma^2,
\ee
relations which we generalize and test empirically in the following.  From the data presented in {\sc mrr}, one observes 
that $\phi$ was rather large on the {\sc nyse} in 1990: $\phi/\lambda_\infty {\cal R}_1 \sim 1-2$.

Note that in the simplest case of independent trade signs ($\rho=0$), the impact function is time independent. 
In the absence of extra compensation for the market makers, $\phi=0$ and the above 
equation reduces to ${\cal R}_1 =  S/2$. In economical terms, this last equality has a very simple meaning: it indicates that on average, 
the new mid-price after the transaction $m_{i+1}=m_i+\epsilon_i {\cal R}$ is equal to the last transaction price $m_i+ \epsilon_i S/2$, 
and therefore that ${\cal R}_1 =  S/2$ is precisely the condition where both market orders and limit orders have zero ex-post cost. This
is more generally the meaning of the {\sc mrr} relation, Eq. (\ref{spreadmrr}): the transaction price is exactly equal to the 
expected long term value of the mid-point. 

It is interesting to discuss the cost of limit orders ${\cal C}_L$ slightly differently. Suppose one wants to trade at a random 
instant in time. Compared to the initial mid-point value, the average execution cost of an infinitesimal buy limit order is given by:
\be\label{limitmrr}
{\cal C}_L = \frac{1}{2} \left(-\frac{S}{2}\right) + \frac{1}{2} \left({\cal R}_1+{\cal C}_L^+ \right):
\ee
with probability $1/2$, the order is executed right away, $S/2$ below the mid-point; otherwise, the mid-point moves on average
by a quantity ${\cal R}_1$, to which must be added the cost of a limit order conditioned to the last trade being a buy, 
${\cal C}_L^+$, for which a similar equation can be obtained:
\be
{\cal C}_L^+ = \frac{1-\rho}{2} \left(-\frac{S}{2}\right) + \frac{1+\rho}{2} \left({\cal R}_1^+ +{\cal C}_L^{++}\right),
\ee
with obvious notations. Since the {\sc mrr} model is Markovian, one has ${\cal R}_1^+={\cal R}_1$ and ${\cal C}_L^{++}=
{\cal C}_L^{+}$, so that:
\be
{\cal C}_L^+ = -\frac{S}{2} +  \frac{1+\rho}{1-\rho} {\cal R}_1.
\ee
Plugging this last relation in Eq. (\ref{limitmrr}), we finally find:
\be
{\cal C}_L = -\frac{S}{2} +  \frac{2}{1-\rho} {\cal R}_1.
\ee
Imposing that ${\cal C}_L\equiv 0$, one recovers the {\sc mrr} relation between the spread and the asymptotic impact (Eq. 
(\ref{spreadmrr}) with $\phi=0$). Note, however, the cost of a market order {\it compared to the initial mid-point value} 
is $S/2$ within the {\sc mrr} model -- but of course the order is still executed at the `right' long term value of the stock. 

\subsection{Real markets are more complicated}

The above model, although suggestive and capturing the essence of the correlation between spread, impact and volatility, is however
not fully satisfactory since it completely neglects the very broad distribution of traded volumes (often found to be log-normal, or 
power-law tailed) and, more importantly, the non-Markovian, long ranged correlation of the trade signs, which is found to decay as 
\cite{QF04,Farmer1}:\footnote{These long ranged correlations were also noted in e.g. \cite{Chordia2,Hopman}, 
 but the detailed shape of the tail of $C(\ell)$ was not investigated, and its long-memory nature not discussed.} 
\be
C(\ell) = \langle \epsilon_i \epsilon_{i+\ell} \rangle \approx \frac{c_0}{\ell^\gamma}, \quad \gamma < 1
\ee
instead of the fast, exponential decay assumed in the {\sc mrr} model. Because the exponent $\gamma$ is found to be less than unity, 
the correlation function is not integrable, which technically makes the series of trade signs a long-memory process. As emphasized 
in \cite{QF04,QF05}, this imposes a number of non-trivial constraints on price impact for the returns to remain uncorrelated while 
the order flow is strongly auto-correlated. In particular, simple models (such as Huang and Stoll's \cite{HS,Stoll}) where price changes 
include a term proportional to $\epsilon_i$ would lead to strong super-diffusion (trends) of prices on the long run \cite{QF04}, in 
disagreement with empirical data.

The volume-dependent lagged impact is now defined as:\footnote{In the definition of ${\cal R}$, care has been taken to remove
any long term trend of the mid-point. In any case, since $\langle \epsilon \rangle$ is close to zero, this trend contribution 
would very nearly vanish.} 
\be
{\cal R}_\ell(v) = \langle \left. \epsilon_i \cdot (m_{\ell+i}-m_i)\rangle\right|_{v_i=v}.
\ee 
In the {\sc mrr} model, $v$ takes a single value and ${\cal R}_\ell(v)$ reduces to the previously defined quantity. 
The function ${\cal R}_\ell(v)$ was studied in detail in \cite{QF04}. To a good level of approximation, the 
following factorization property is found to hold: ${\cal R}_\ell(v) \approx R(\ell) f(v)$, where $f(v)$ is a strongly concave 
function, and $R(\ell)$ an increasing function of $\ell$ that varies by a factor of $\sim 2$ when $\ell$ increases 
from $1$ to several thousands (corresponding to a few days of trading).\footnote{The true asymptotic behaviour of $R(\ell)$ for 
longer horizons is difficult to determine empirically due to statistical noise, and might in fact be stock dependent, see \cite{QF05} 
for a discussion of this point.} The shape of $R(\ell)$, averaged over a collection of different stocks of the {\sc pse}, is shown 
in Fig. 1, and compared with the simple form assumed in the {\sc mrr} model (see caption for more details). Perhaps more 
importantly, the enhancement factor $\lambda_\infty$ is found empirically to be substantially larger than predicted by 
Eq. (\ref{enhance}). For example, on the pool of 68 {\sc pse} stocks studied below, we find, averaged over all stocks, 
$\lambda_\infty \approx 1.75$ whereas $1/(1-C_1) \approx 1.32$ (see Table 2 of Appendix 2). The difference between the two 
will turn out to play a crucial role in the following.

\begin{figure}[htbp]
\begin{center}
\rotatebox{270}{\resizebox{7.0cm}{!}{\includegraphics{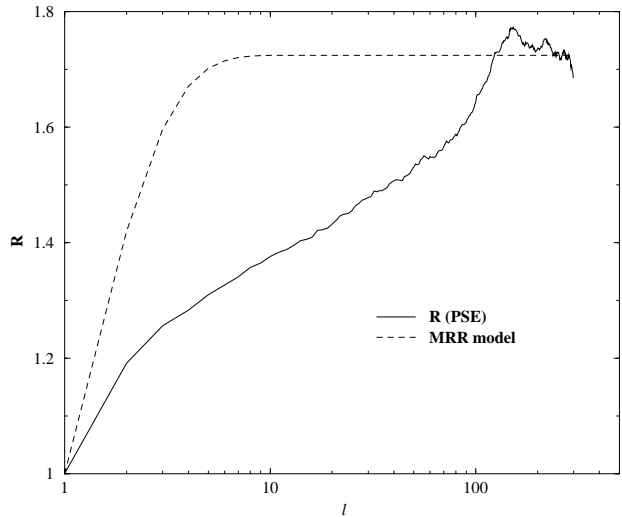}}}
\caption{\small{Average over 68 {\sc pse} stocks of the impact function $R(\ell)$ as a function of $\ell$ (plain line). The average is 
performed by rescaling the individual $R(\ell)$ such that $R(\ell=1) \equiv 1$, and by rescaling 
$\ell$ by the average daily number of trades and multiplying by 100. Dotted line: prediction of the {\sc mrr} model with $\rho=3/7$, such that $\lambda_\infty=1.75$. 
The discrepancy with empirical data shows the importance of correctly accounting for long-range correlations in order flow.
}}
\label{avR}
\end{center}
\end{figure}

\subsection{Market order strategies}

In this section and below, we want to show how the simple relations derived in the {\sc mrr} model can be extended and 
tested in the general case of fluctuating volumes and long-ranged correlation of trade signs. A first idea is to measure empirically 
the average execution cost of market-orders. One can define the {\it ex-post} cost ${\cal C}_M(T)$ the difference between the 
transaction price at (trade-)time $i$ and the mid-point price at time $i+T$ later, with $T \gg 1$ but still much smaller than the 
typical horizon of the trading strategy itself (a few days or more), in order not to mix in the quality of the decision to trade. 
The above definition of execution cost marks the trade to market 
after $T$ and is referred to the {\it realized spread} in the literature \cite{Bessembinder02,Stoll}. The volume weighted averaged 
cost (over $N$ trades) of a single market order over horizon $T$ is therefore:\footnote{Note that this definition neglects the fact that one single large 
market order may trigger transactions at several different prices, up the order book ladder, and pay more than the nominal spread. 
Nevertheless this situation is empirically quite rare on the markets we are concerned with, and 
corresponds to only a few percents of all cases \cite{Farmer5}.}
\be\label{costMO}
{\cal C}_M(T) = \frac{1}{N \langle v \rangle} \sum_{i=1}^N  \epsilon_i v_i (m_i + \epsilon_i \frac{S_i}{2} - m_{i+T}) \equiv
\frac{\langle v S \rangle}{2\langle v \rangle} - \frac{\langle v {\cal R}_T(v) \rangle}{\langle v \rangle}.
\ee
The choice $T \gg 1$ allows us to use the asymptotic value of $R$, $R(\ell \gg 1) \approx
\lambda_\infty R(1)$, where we have introduced a factor $\lambda_\infty$ in conformity with the notation of the previous section. 
Using the factorization property of ${\cal R}_\ell(v)$, we finally obtain for the average cost of a single market order:
\be\label{cc}
{\cal C}_M(T \gg 1) = \frac{\langle v S \rangle}{2\langle v \rangle} - \lambda_\infty \frac{\langle v {\cal R}_1(v) \rangle}{\langle v \rangle},
\ee
meaning, as intuitively clear, that this cost is positive when spreads are large, but may become negative if the total price 
impact $\lambda_\infty {\cal R}_1$ is large. In the plane $x={\langle v {\cal R}_1(v) \rangle}/{\langle v \rangle}$, $y=
{\langle v S \rangle}/{\langle v \rangle}$ (which will repeatedly be used below to represent empirical data) 
the condition ${\cal C}(T \gg 1)=0$ defines a straight line of slope $2 \lambda_\infty$ separating an upper region where market orders 
are on average costly, from a region where single market orders are favored: see Fig. 2. 

The above computation suggests an upper bound on the spread, which we establish more rigorously in the next section. 
For larger spreads, the positive average cost of market order would deter 
their use; limit orders would then pile up and reduce the spread. What would happen if the spread was below the red line 
of slope $2 \lambda_\infty$ in Fig. 2? Naively, market orders have a negative cost in that region, and 
one might be able to devise profitable strategies based solely on market orders. The idea would be to try 
to benefit from the impact term ${\cal R}_\infty$ in the 
above balance equation. The growth of ${\cal R}_\ell$ ultimately comes from the correlation between 
trades, i.e. the succession of buy (sell) trades that typically follow a given buy (sell) market order.  
The simplest `copy-cat' strategy which one can rigorously test on empirical data is to place a market order with 
vanishing volume fraction 
(not to affect the subsequent history of quotes and trades), immediately following another market order. 
This strategy suffers on 
average from the impact of the initial trade, used as a guide to guess the direction of the market. 
Therefore, the profit ${\cal G}_{CC}$ of such a copy-cat strategy, marked to market after a long time and
neglecting further unwinding costs, is reduced to:\footnote{A more rigorous estimate of the gain of a copy-cat 
strategy participating to all the trades can be obtained following the method outlined in the next section.}
\be\label{cc3}
{\cal G}_{CC} = [\lambda_\infty-1] \frac{\langle v {\cal R}_1(v) \rangle}{\langle v \rangle}
-\frac{\langle v S \rangle}{2\langle v \rangle}.
\ee
Imposing that this gain is non-positive, one obtains a lower line in the plane $x,y$, of slope $2 (\lambda_\infty-1)$. 
Only below this green line can the above infinitesimal copy-cat strategy be profitable. 
We therefore expect markets to operate above this line and below the red line of slope $2\lambda_\infty$.
Note however that market orders below the $2 \lambda_\infty$ line are not necessarily favorable in practice, since 
the cost for executing a {\it series} of market orders (which is the typical situation faced by large investors, since the
outstanding liquidity is, as noted above, always quite small) must include the impact of past trades and 
this increases their average cost. Hence, the slope of the effective zero-cost line for a series of market orders is indeed 
smaller than $2\lambda_\infty$. Similarly, the long-time impact of an isolated market order, uncorrelated 
with the order flow, is in fact very small \cite{QF04}. These isolated market orders thus also have a positive 
cost, equal to half the spread. The only way to benefit from the average impact ${\cal R}_\ell$ is to 
free-ride on a wave of orders launched by others, as in the above copy-cat strategy. Let us now take the 
complementary point of view of limit orders and determine the region of profitable market making strategies.

\begin{figure}[htbp]
\begin{center}
\rotatebox{270}{\resizebox{7.0cm}{!}{\includegraphics{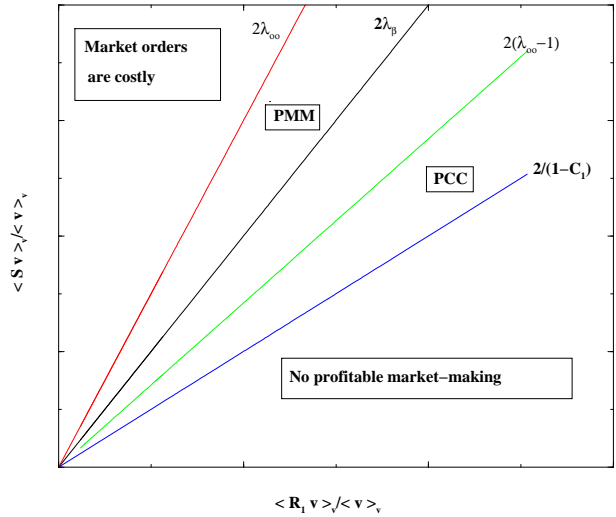}}}
\caption{\small{General ``phase diagram" in the plane $x={\langle v {\cal R}_1(v) \rangle}/{\langle v \rangle}$, 
$y={\langle v S \rangle}/{\langle v \rangle}$, showing several regions: (i) above the red line of slope $2 \lambda_\infty$, 
market orders are costly (on average) and market making is profitable; (ii) below the blue line of slope $\approx 2/(1-C_1)$, limit orders
are costly and no market-making strategy is profitable; (iii) above the black line of slope $2 {\overline \lambda}_\beta$, market making on time scale 
$T$ (or faster) is profitable (PMM); (iv) below the green line of slope $2(\lambda_\infty-1)$, copy-cat strategies can be 
profitable (PCC). Since neither market orders nor liquidity providing should be systematically penalized for markets to ensure steady 
trading, we expect that markets should operate in the `neutral wedge' in between the blue and the red line. Competition between 
liquidity providers should push the market towards the blue line. Since copy-cat strategies should not be profitable either, the PCC green line cannot
lie above this blue line. Note that the blue, red and black lines all coincide within the {\sc mrr} model.
}}
\label{Diag}
\end{center}
\end{figure}

\subsection{An infinitesimal market making strategy}

Our aim is to discuss the profitability of providing liquidity to the market formalizing the idea of infinitesimal 
strategies used in the previous section. 
To do so we compute the gain of a simple market making strategy which consists in participating to a vanishing fraction of 
all trades through limit orders. The simplest strategy is to consider a market maker with a certain time horizon $T$ 
who provides an infinitesimal fraction $\varphi$ of the total available liquidity. As illustrated by Eq. (\ref{costMO}), 
the cost incurred by the market maker comes from market impact: the price move between $0$ and $T$ is anti-correlated with 
the accumulated position. When the crowd buys, the price goes up while the market making strategy accumulates a short position which would be costly 
to buy back at time $T$, and vice-versa. More precisely, we consider a {\it steady-state} market making strategy 
(which avoids explicit unwinding costs). The strategy is such that volume offered dynamically depends on the accumulated position, 
which insures that the inventory is always bounded. We choose the tendered fraction $\varphi$ to be given by: 
$\varphi_i= \varphi_0(1 + \alpha V_i \epsilon)$, where $V_i$ is the (signed) position accumulated up to 
time $i^-$, and $\epsilon=+1$ for orders placed at the ask and $\epsilon=-1$ for orders placed at the bid. 
This mean-reverting strategy insures that the typical position is always bounded. One can now use this strategy for
an arbitrary long time $T$; its profit \& loss is simply given by: 
\be
{\cal G}_L = \sum_{i=0}^{T-1} \varphi_i \epsilon_i v_i (m_i +\epsilon_i \frac{S_i}{2}).
\ee
For large $T$, one can replace this expression by:
\be
{\cal G}_L = T \langle \varphi_i \epsilon_i v_i (m_i +\epsilon_i \frac{S_i}{2}) \rangle
\ee
with $O(T^0)$ corrections due to the residual position at $T$. Discarding the constraint $\varphi_i \geq 0$ and 
neglecting volume-volume correlations, which are much smaller than sign-sign correlations \cite{QF04,Farmer1}, we finally find:
\be\label{gainbeta}
\frac{{\cal G}_L(\beta)}{T \varphi_0 \langle v \rangle} = \frac{\langle v S \rangle}{2\langle v \rangle} \left[1 - \frac{1-\beta}{\beta}
\sum_{\ell=1}^\infty \beta^\ell C(\ell) \right] -  \frac{1-\beta}{\beta}
\sum_{\ell=1}^\infty \beta^\ell  \frac{\langle v {\cal R}_\ell(v) \rangle}{\langle v \rangle},
\ee
where $\beta =1- \alpha \varphi_0 \langle v \rangle$ fixes the typical time scale of the market making strategy. The above expression
is exact in the limit $\alpha \to 0$, and only approximate otherwise. When $\beta \to 0$
(fast market making), Eq.(\ref{gainbeta}) reduces to:
\be
\frac{{\cal G}_L(\beta \to 0)}{T \varphi_0 \langle v \rangle} \approx \frac{\langle v S \rangle}{2\langle v \rangle} \left[1 - C_1 \right] - 
\frac{\langle v {\cal R}_1(v) \rangle}{\langle v \rangle},
\ee
whereas $\beta \to 1$, corresponding to slow market making, yields:
\be
\frac{{\cal G}_L(\beta \to 1)}{T \varphi_0 \langle v \rangle} = \frac{\langle v S \rangle}{2\langle v \rangle}  - 
\frac{\langle v {\cal R}_\infty(v) \rangle}{\langle v \rangle}.
\ee
Setting ${\cal G}_L(\beta)$ to zero leads to a linear relation between spread and impact:
\be
\frac{\langle v S \rangle}{\langle v \rangle}  
= 2 {\overline \lambda}_\beta \frac{\langle v {\cal R}_1(v) \rangle}{\langle v \rangle}.
\ee
Using the empirical shape of ${\cal R}_\ell$ and $C(\ell)$, the slope $2 {\overline \lambda}_\beta$ 
is found to increase between $\approx 2/(1-C_1)$ and $2 \lambda_\infty$ when $\beta$ increases. Contrarily to 
market orders which benefit from the growth of the impact ${\cal R}_\ell$ with time, slow market making 
is suboptimal. When $\beta \to  1$, ${\overline \lambda}_\beta \to \lambda_\infty$ and 
the lower limit of profitability of very slow market making is precisely the red line of Fig. 2 where 
market orders become profitable. Faster strategies correspond 
to smaller values of ${\overline \lambda}_\beta$, closer to $1/(1-C_1)$, leading to an extended region 
of profitability for market making. From the assumption that the above market making strategy for any 
value of $\beta$ should be at best marginally profitable (since one might find more sophisticated strategies, which take 
full advantage of the correlations between signs and volumes), we finally obtain the following bound 
between spread and impact: 
\be
\label{mmmg}
\frac{\langle v S \rangle}{\langle v \rangle}  \leq  \frac{2}{1-C_1} \frac{\langle v {\cal R}_1(v)\rangle}{\langle v \rangle},
\ee
defining the blue line of slope $2/(1-C_1)$ in the $x,y$ plane of Fig. 2. Consistently with the {\sc mrr} model, when 
$\lambda_\infty = 1/(1-C_1)$, the blue and red line of Fig. 2 exactly coincide. Using that fact that 
${\cal R}_1^{n+} \leq {\cal R}_1^{(n-1)+}$, a simple generalisation of the argument presented at the end of Sect. 2.1 allows one 
to show that the cost of limit orders is indeed negative above the blue line.

\subsection{Theoretical analysis: conclusions}

Eqs.~(\ref{cc},\ref{cc3},\ref{mmmg}) and the resulting microstructural ``phase diagram'' of Fig. 2 are 
our central results. 
These equations show that the cost or profitability of infinitesimal market and limit order strategies
can be estimated 
from empirical data alone, without having to make any further assumption on the fraction of 
informed trades, the correlation between trades, etc. In order to proceed, we made two approximations.
Firstly, we assumed that these strategies could be made infinitesimal, 
which allows us to neglect their impact on the price dynamics. In practice, trades occur in discrete volume, and 
strictly speaking the assumption of infinitely small volumes does not hold. However, the volume of typical trades is 
much larger than the minimum size, which suggests that this approximation is accurate.
Secondly, we neglected all direct transaction costs, which obviously affect profitability. These costs are in general very small compared 
to the spread, and can therefore also reasonably be neglected.

Our main result is that profitability, perhaps surprisingly, depends on the {\it frequency} of these strategies, 
a result closely related to the anomalous time dependence of the impact function. Market orders are favored at low frequencies, 
when impact has fully developed, whereas limit orders are favored at high frequencies, where impact is still limited and the execution
probability significant. 

Our analysis delineates, in the impact-spread plane, a central
wedge bounded from above by a slope $2 \lambda_\infty$ and from below by a slope $\approx 2/(1-C_1)$, within which both market 
orders and limit orders are viable. In the upper wedge, market orders would always be costly and would be substituted 
by limit orders. In the lower wedge, market making strategies, even at high frequencies, would never eke out any profit. 
Such a market would not be sustainable in the absence of any incentive to provide liquidity. But if the spread happened to fall in this
region, the enhanced flow of market orders would soon reopen the gap between bid and ask.

Our next assumption is that simple statistical strategies must have negative or marginal profit.
These is quite reasonable since high-frequency strategies carry relatively small risks. Applying this
idea to market making strategies, we conclude that competition between liquidity providers
will push the spreads close to the lower limit, corresponding to the blue line of slope $\approx 2/(1-C_1)$ in Fig. 2. 
Now, since market taking (copy-cat) strategies should not be profitable either, the green line 
of slope $2(\lambda_\infty-1)$ should necessarily lie below the blue line, leading to the following 
inequality on the asymptotic impact enhancement factor $\lambda_\infty$:
\be
1 \leq \lambda_\infty \leq 1+\frac{1}{1-C_1},
\ee
where the lower bound comes from the existence of correlation between trades (see Eq. (\ref{enhance})). 
In other words, the impact function cannot grow more than roughly twice its initial value, otherwise 
statistical arbitrage would set in. Interestingly, our data is compatible with the above bound;  in practice the blue and green 
lines turn out to be not very far from each other. 

Finally, we note that market microstructure studies insist on large inventory risks being an 
important determinant of the 
bid-ask spread. However, large inventories correspond to long horizons and slow market making. 
Our analysis above shows that accumulating inventories on a long horizon is not only risky, but may also 
be extremely costly on average. When $\lambda_\infty > 1/(1-C_1)$, 
market making on large horizons is significantly more costly than on short horizons, by an amount 
proportional to the spread itself. This is a very strong effect, which makes the existence of 
low-frequency market makers very unlikely. Therefore, inventory risk by itself should not be important 
in determining the value of the spread, at least on electronic markets. 

In conclusion, we expect that electronic markets should operate in the vicinity of the blue line of Fig. 2, imposing a 
linear relation between spread and market impact of slope close to $2/(1-C_1)$. This is what we test on empirical data 
in the following section.

\section{Comparison with empirical data}

\subsection{Small tick electronic markets}

We first consider small tick electronic markets, such as the Paris Stock Exchange ({\sc pse}) or Index Futures. 
The case of large tick stocks is different since in this case the spread is (nearly) always one tick, with huge volumes 
at both the bid and the ask. The case of such markets will be considered below. 

We studied extensively the set of the 68 most liquid stocks of the {\sc pse} during the year 2002. The 
summary statistics describing these stocks is given in Appendix 2. From the Trades and Quotes data, one has access the 
the bid-ask just before each trade, from which one can obtain the sign and the volume of each trade (depending on whether the trade 
happened at the ask or at the bid) and the mid-point just before the trade. From this information, one computes 
the quantities of interest, such as the instantaneous impact function ${\cal R}_1$, the one-lag correlation $C_1$, 
the spread $S$ and $\lambda_\infty$. 
Note that we have
removed `block trades', which appear as transactions with volumes larger than what is available at the best price that 
are not followed by a change of quotes. Clearly, these block trades are outside the scope of the above arguments; in any case they
represent typically a $5 - 10 \%$ fraction of the total number of trades and do not significantly affect the following results. 

We test the above ideas in two different ways -- for a given stock across time, and across all
different stocks. Since both the spread and impact vary with time, one can measure `instantaneous' 
quantities by averaging for a given stock $\langle S v\rangle/\langle v\rangle$ and 
$\langle v {\cal R}_1(v)\rangle/\langle v\rangle$ over a number of successive trades. In the example 
of Fig. 3, each point corresponds to an average over 10000 non overlapping trades, corresponding to 2 days
of trading in the case of France Telecom in 2002.  Doing so we obtain quantities that vary by a factor 5 that allows us 
to test the linear dependence predicted by Eqs.(\ref{cc3},\ref{mmmg}). For France Telecom, we find that $\lambda_\infty$ is 
close to the average value $1.85$ shown in Fig. 1. Therefore $2(\lambda_\infty-1) {\stackrel{<}\approx} \, 2$ in this case, meaning that 
copy-cat market making strategies are impossible, as expected for highly liquid stocks. We also find that $C_1 \approx 0.14$ (see
Appendix 2). Our results shown in Fig.\ref{FT} are in good agreement 
with the above theoretical bounds, even for averages over rather short time scales. A linear fit with zero intercept gives a slope 
equal to $2.14$, to be compared with $2/(1-C_1) \approx 2.32$, meaning that providing liquidity is hardly rewarded at all for this very liquid, small tick stock. 
In fact, if the intercept of the linear fit is left free, its value (which should equal the `processing costs' $2 \phi$ 
in the {\sc mrr} model) is found to be slightly negative.

\begin{figure}[htbp]
\begin{center}
\rotatebox{270}{\resizebox{7.0cm}{!}{\includegraphics{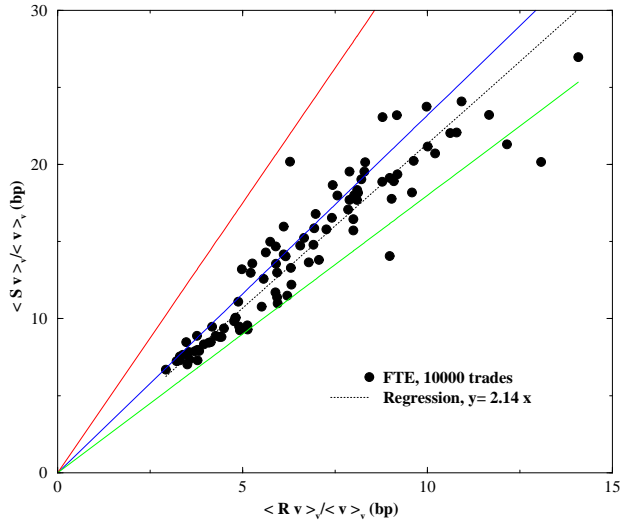}}}
\caption{\small{France Telecom in 2002. Each point corresponds to a pair 
($y=\langle v S \rangle/\langle v\rangle$, $x=\langle v {\cal R}_1 \rangle/\langle v\rangle$), computed by averaging 
over 10000 non overlapping trades ($\sim$ two trading days). Both quantities are expressed in basis points. 
We also show the different bounds, Eqs. (\ref{cc},\ref{cc3},\ref{mmmg}), and a linear fit that gives a slope of $2.14$. 
The correlation is $R^2=0.93$.}}
\label{FT}
\end{center}
\end{figure}

We also test Eq.(\ref{mmmg}) cross-sectionally in Fig. 4, using the above 68 different stocks of the {\sc pse}.
The relative values of the spread and the average impact also varies by a factor 5 between the different stocks, 
which enables to test the linear relations (\ref{cc3},\ref{mmmg}). Once again we find a good agreement with the predicted 
bound, and the linear fit with zero intercept gives a slope of $2.86$, while $\langle 2/(1-C_1) \rangle \approx 2.64$.
Hence, fast market making strategies are on average weakly profitable on the {\sc pse}. 
However, the intercept of a two-parameter regression is very slightly negative, showing that no order processing costs component 
can be detected on these fully electronic markets.

\begin{figure}[htbp]
\begin{center}
\rotatebox{270}{\resizebox{7.0cm}{!}{\includegraphics{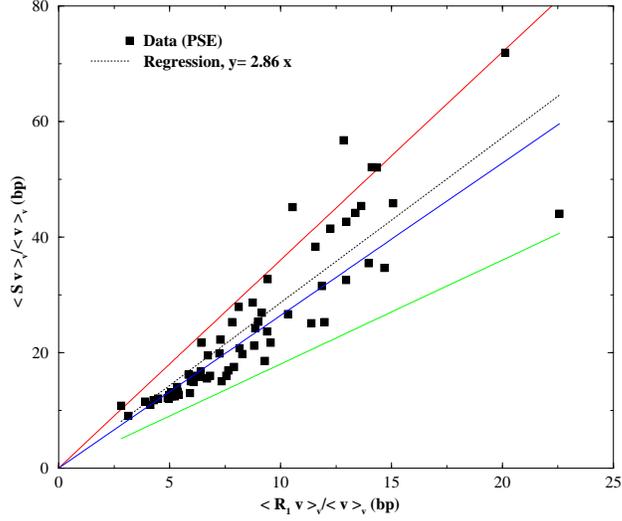}}}
\caption{\small{68 stocks of the Paris Stock Exchange in 2002. Each point corresponds to a pair 
($y=\langle v S \rangle/\langle v\rangle$, $x=\langle v {\cal R}_1 \rangle/\langle v\rangle$), computed by averaging 
over the year. Both quantities are expressed in basis points. We also show the different bounds, 
Eqs. (\ref{cc},\ref{cc3},\ref{mmmg}), 
and a linear fit that gives a slope of $2.86$, while $\langle 2/(1-C_1) \rangle \approx 2.64$. The correlation is $R^2=0.90$.}}
\label{cac}
\end{center}
\end{figure}

\begin{figure}[htbp]
\begin{center}
\rotatebox{270}{\resizebox{7.0cm}{!}{\includegraphics{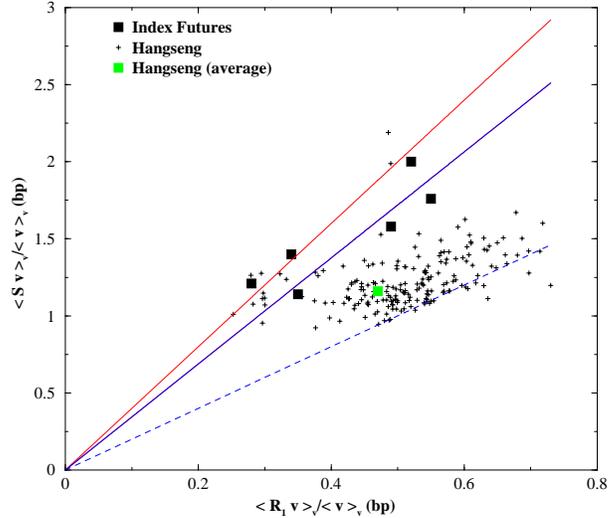}}}
\caption{\small{Small tick Index Futures in 2005: {\sc cac}, {\sc dax}, {\sc ftse}, 
{\sc ibex}, {\sc mib}, {\sc smi}, {\sc hangseng}. Each black square corresponds to a pair 
($y=\langle v S \rangle/\langle v\rangle$, $x=\langle {\cal R}_1 v\rangle/\langle v\rangle$), computed by averaging 
over the year, while small crosses are computed by averaging over 1000 non overlapping trades on the {\sc hangseng}
futures. Both quantities are expressed in basis points. We also show the bounds, Eqs. (\ref{mmmg},\ref{cc}), with $1/(1-C_1)\approx 1$
(dotted blue line), corresponding to the {\sc hangseng}, and $1/(1-C_1)\approx 1.72$ (full blue line), 
corresponding to the average over all other
futures.}}
\label{FUT}
\end{center}
\end{figure}

It is also interesting to analyze small tick Futures markets, for which the typical spread is ten times smaller than on 
stock markets. We have studied a series of small tick Index Futures in 2005 (except the {\sc mib} for which the data is 
2004), again both as a function of time and
across the 7 indexes of our set. For most contracts, the value of $C_1$ is quite large ($\langle C_1 \rangle
 \approx 0.42$) except for the {\sc hangseng} where $C_1 \approx 0.035$. 
Results are shown in Fig. \ref{FUT}; the bounds are again quite well obeyed both across 
contracts and across time, even when the time averaging is restricted to only 1000 consecutive trades. 
This shows that on these highly liquid contracts, where the transaction rate as high as a few per second, 
the equilibrium between spread and impact is reached very quickly.


\subsection{NYSE stocks}

The case of the {\sc nyse} is quite interesting since the market is still ruled by specialists, who however 
compete to provide liquidity with other market participants placing limit orders. We again test Eqs.(\ref{cc},\ref{mmmg}) cross 
sectionally, using the set of the 155 most actively traded stocks on the {\sc nyse} in 2005.\footnote{The list of the
155 names is available on request.} We use the quoted bid-ask posted by the specialist. We have first determined the average 
impact function $R(\ell)$, which has a shape roughly similar to Fig. 1, although the asymptotic plateau value is slightly larger, 
leading to $\lambda_\infty \approx 2.1$. On the other hand, $1/(1-C_1)$ is also slightly larger, 
equal to $1.39$.

Plotting the data in the spread-impact plane, we now find that the empirical results cluster around to the upper red line 
limit where market orders become costly. The regression has a significantly larger slope of $3.3$ 
and now a positive intercept 
$2 \phi \approx 1.3$
basis points.\footnote{This is five times smaller than the average spread, leading to $\phi/\theta \sim 0.25$, much smaller
than the result $\phi/\theta \sim 1 - 2$ found within {\sc mrr} model in 1990, or a similar value reported in \cite{Stoll}.}
This suggests that, perhaps not surprisingly, the existence of monopoly rents on {\sc nyse}: market makers post spreads that are 
systematically over-estimated compared to the situation in electronic markets, with a non-zero extrapolated spread $2 \phi$ for 
zero market impact. 
This result is in agreement with the study of Harris and Hasbrouck performed
in the early 90's on the {\sc nyse} \cite{HH}, which showed that limit orders were more favorable than market orders, 
and also with Handa and Schwartz \cite{HandaS}, who showed that pure limit order strategies were indeed profitable. On the other hand, 
the value of the regression slope on the purely electronic {\sc pse} show that pure limit order strategies can only be marginally 
profitable. We have checked that using the traded spread instead of the quoted spread does not change 
appreciably the above conclusions.

\begin{figure}[htbp]
\begin{center}
\rotatebox{270}{\resizebox{7.0cm}{!}{\includegraphics{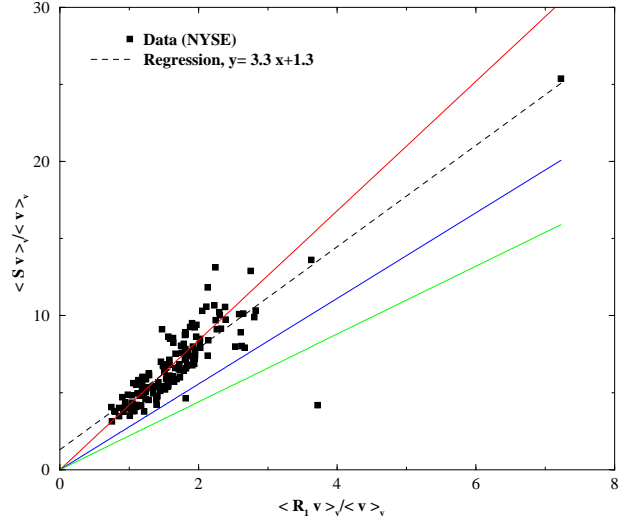}}}
\caption{\small{155 stocks of the {\sc nyse} 2005. Each point corresponds to a pair 
($y=\langle v S \rangle/\langle v\rangle$, $x=\langle v {\cal R}_1 \rangle/\langle v\rangle$), computed by averaging 
over the year. Both quantities are expressed in basis points. We also show our bounds, Eqs. (\ref{cc},\ref{cc3},\ref{mmmg}). 
The data shows clearly that market orders are less favorable than in the electronic Paris Bourse. The regression now has a 
positive intercept of $1.3$ bp with an $R^2=0.87$.}}
\label{nyser1}
\end{center}
\end{figure}

\subsection{The case of large tick electronic markets}

{\it A priori}, the string of arguments leading to Eq. (\ref{mmmg}) does not directly apply in the case where the tick size is large. 
In that case the spread $S$ is most of the time stuck to its minimum value, i.e. one tick, while the size of the queue $q$ 
at the bid and at the ask tends to be extremely large (see e.g. Appendix 2, Table 3). Because of the large value of 
the spread, limit orders appear to be favorable, but huge limit order volumes accumulate 
as liquidity providers attempt to take advantage of the spread. The size of the queue $q$ at the bid or at the ask is thus 
much larger than the typical value of the traded 
volume at each transaction $v$: $v/q \approx 0.01$ (see Table 3), to be compared with $v/q \approx 0.2 - 0.3$ 
(see Appendix 2, Table 2) for smaller
tick stocks. Therefore, the simple market making strategy considered above, which assumes that one can participate to a 
small fraction of {\it all} transactions, cannot be implemented. We thus expect that the spread on these markets will
be substantially larger than predicted by the bound Eq. (\ref{mmmg}), because the competition between liquidity providers, 
that acts to reduce the spread, cannot fully operate. We indeed find that the ratio between $\langle v S \rangle$ and 
$\langle v {\cal R}_1 \rangle$ is large for large tick stocks. For example, in the case of Ericsson, during the 
period March-November 2004, for which the tick size is $\sim 50$ bp, we find  
$\langle v S \rangle/\langle v {\cal R}_1 \rangle \approx 4.5$. However, we also find on the same data 
that $\lambda_\infty \approx 4.5 \pm 1.$, meaning that market orders are in fact not systematically unfavored in these large tick 
electronic markets. In fact, all data points are found to lie between our bounds, Eqs. (\ref{cc},\ref{mmmg}), but indeed 
significantly higher than the blue line of Fig. 2 in this case.

\subsection{Comparison with empirical data: conclusion}

Our empirical analysis shows that on liquid markets, an approximate symmetry between limit and market 
orders indeed holds, in the sense that neither market orders nor limit orders are systematically unfavorable. Markets 
operate in the `neutral wedge' of Fig. 2. 

For fully electronic markets, competition for providing liquidity is efficient in keeping 
the spread close to its lowest value, marginally compensating impact cost. There is therefore hardly any room for market making
strategies. Although the cost of {\it isolated} market orders is found to be negative, the empirically established proximity of the 
blue and green line in Fig. 2 means that there is no room for simple market {\it taking} strategies either. In this discussion, 
time horizon and long range correlations in the order flow play an important role, overlooked in previous studies \cite{HS,MadS,Stoll}: 
somewhat paradoxically, liquidity providers as a whole offer average negative costs to market orders but high frequency 
market making strategies still manage to get (marginally) compensated. Our analysis shows that the ecology between liquidity 
takers and liquidity providers turns out to be considerably more complex than anticipated by Handa \& Schwartz \cite{Handa}:
when costs are computed on large time scale, limit orders are in average costly.  This implies that a significant fraction 
of limit orders cannot be due to market makers, since limit orders as a whole are in arrears. The common assumption that limit
orders can be attributed to liquidity providers compensated by the spread cannot be correct in electronic markets. This 
argument can only concern a small fraction of high-frequency market makers, whose existence is nevertheless crucial to 
prevent liquidity crises.

On the {\sc nyse}, spreads appears to be significantly larger: isolated market orders are now marginally costly. 
A linear relation between spread and impact still applies, albeit with a larger slope and 
a residual intercept, corresponding to market maker monopoly rents, which are absent in electronic markets.

\section{Liquidity {\it vs.} volatility}
 
 \subsection{Theoretical considerations}

 Consider again the {\sc mrr} model discussed above, which predicts a simple relation between volatility and 
 impact, Eq. (\ref{sigmamrr}). Using the relation between spread and impact established above, this suggests a direct
 link between volatility per trade and spread, which we motivate and test in this section. 
 
\begin{figure}[htbp]
 \begin{center}
 \rotatebox{270}{\resizebox{7.0cm}{!}{\includegraphics{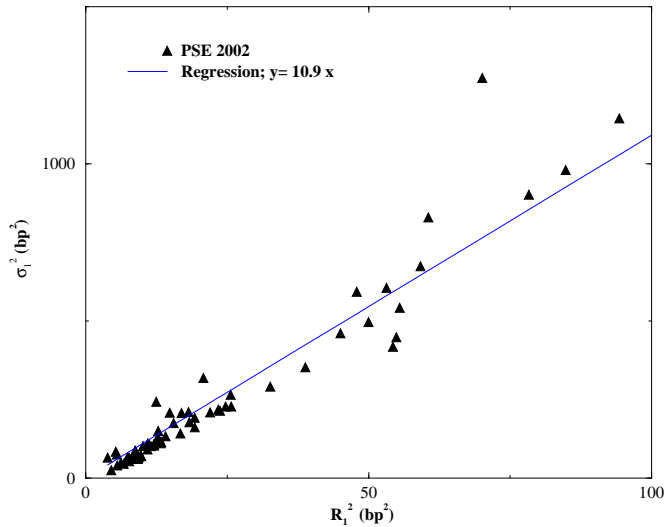}}}
 \caption{\small{Plot of $\sigma_1^2$ vs.  ${\overline{\cal R}}_1^2$, showing that the linear relation Eq. (\ref{dd2}) holds quite 
 precisely with $\Sigma^2 = 0$ and $a \approx 10.9$. (The intercept of the
 best affine regression is even found to be slightly negative). Data here corresponds to the 68 stocks of the {\sc pse} in 2002. 
 The correlation is very high: $R^2=0.96$.
 }}
 \label{sigvsR}
 \end{center}
 \end{figure}

 By definition of the volatility per trade $\sigma_1^2 = \langle (m_{\ell+1}-m_\ell)^2\rangle$ and of the
 instantaneous impact $r_{1,i} \equiv (m_{i+1}-m_i).\epsilon_i$, one has as an identity:
 \be 
 \label{dd}
 \sigma _1^2 \equiv \langle r_{1,i}^2 \rangle.
 \ee
 The instantaneous impact $r_{1,i}$ is expected to fluctuate over time for several reasons. First, 
 the volume of the trade, the volume in the book and the spread strongly fluctuate with time. For example, on the {\sc pse}, 
 the spread has a distribution close to an exponential, hence one has $\langle S^2 \rangle \approx 2 \langle S \rangle^2$
 (see Table 2, Appendix 2).\footnote{The distribution appears to be a power-law on other markets \cite{Mike}, but 
 this is irrelevant for the following discussion.} Large impact fluctuations may also 
 arise from quote revisions due to addition or cancellation of some limit orders. Second, there
 might also be important news affecting the `fundamental price' of the stock. These result in large, instantaneous jumps of 
 the mid-point, unrelated to the trading activity itself. 
 In order to account for both effects, we write, generalizing the above {\sc mrr} relation:
 \be\label{dd2}
 \sigma _1^2 = a {\overline{\cal R}}_1^2 + \Sigma^2,
 \ee
 where ${\overline{\cal R}}_1 \equiv \langle {\cal R}_1(v)\rangle$ is the average impact after one trade, $a$ is a coefficient measuring the variance of impact fluctuations and $\Sigma^2$ is the news component of the volatility (see Section 2.1). 
 A specific model for Eq. (\ref{dd2}) was worked out in \cite{QF04}, and tested on France-Telecom (see also \cite{Rosenow}). 
 Here, we establish that this relation holds quite precisely across different
 stocks of the {\sc pse}, with a correlation of $R^2=0.96$ (see Fig. (\ref{sigvsR})). Perhaps surprisingly, the exogenous 
 `news volatility' contribution
 $\Sigma^2$ is found to be small. (The intercept of the
 best affine regression is even found to be slightly negative). This could be related to the observation made in 
 Farmer et al. \cite{Farmer5} 
 that for most price jumps, some limit orders are canceled to slowly and get `grabbed' by fast market orders, which 
 means that most of these events are already included in ${\overline{\cal R}}_1$, in line with our general statements
 on the approximate symmetry between limit and market orders.\footnote{One could argue that our results simply 
 show that the news volatility $\Sigma$ itself is proportional to ${\overline{\cal R}}_1$ and thus to the spread $S$. 
 However, there is no reason why this should {\it a priori} be the case. For example, a model where jumps of typical amplitude $J$
 have a small probability per trade $p$ leads to $\Sigma = \sqrt{p} J$, whereas the cost of such jumps, 
 contributing to $S$, is $p J \ll \Sigma$.} 
 In the following, we will therefore neglect $\Sigma^2$, 
 as suggested by Fig. (\ref{sigvsR}): in this sense the volatility of the stocks can be mostly attributed to market activity and trade impact.
 This is in agreement with the conclusions of Lyons and Evans on currency markets \cite{Lyons}; see also the 
 discussion in \cite{QF04,Hopman}.

\begin{figure}[htbp]
 \begin{center}
 \rotatebox{270}{\resizebox{7.0cm}{!}{\includegraphics{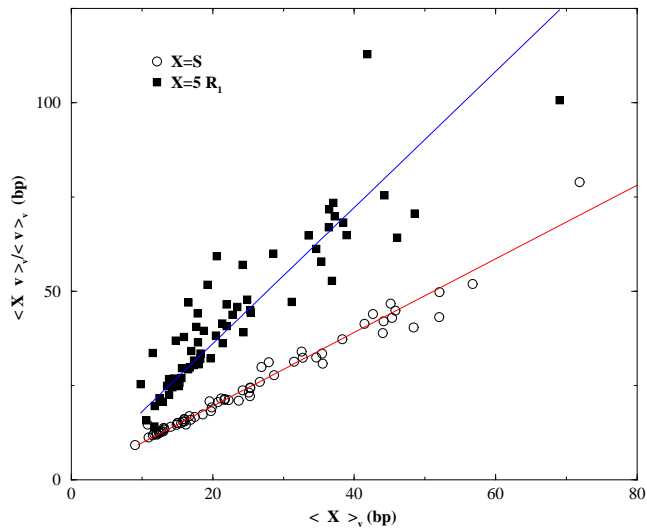}}}
 \caption{\small{Plot of $\langle v X\rangle _v/\langle v \rangle$ vs.  $\langle X\rangle $, where $X$ is either the
 spread $S$ or the instantaneous impact ${\cal R}_1(v)$ (multiplied by a factor $5$ for clarity). The quality of the linear 
 regression tests our universality assumption, which is excellent for $S$ ($R^2=0.98$) and satisfactory for ${\cal R}_1$ ($R^2=0.9$). 
 The value of $b \approx 1.02$ and $b' \approx 1.80$ are given by the slope of these regressions. Data here corresponds to the 68 stocks of the {\sc pse}
 in 2002. 
 }}
 \label{xv}
 \end{center}
 \end{figure}

 Our final assumption is that of {\it universality}, i.e. when the tick size is small enough and the typical number of shares 
 traded is large enough, all stocks within the same market should behave identically up to a rescaling of the average spread and 
 the average volume. In particular we assume that the statistics of (i) the volume of market orders (ii) the spread S and 
 (iii) the impact ${\cal R}$, and the correlations between these quantities are independent on the stock when these quantities
  are normalized by their average value.\footnote{The universality of the shape of the order book was indeed checked to hold 
  rather well in \cite{BMP}.} This universality implies that:
 \be\label{univ1}
 \langle v S\rangle = b \langle v \rangle \langle S\rangle ,
 \ee
 where $b$ is stock independent. Similarly, 
 \be\label{univ2}
 \langle v {\cal R}_1(v)\rangle _v = b' \langle v \rangle {\overline{\cal R}}_1,
 \ee
 where $b'$ is also stock independent. Note that this assumption is consistent with the empirical observation 
 of \cite{Lillo}, where the impact function ${\cal R}_1(v)$ for different US stocks can indeed be rescaled 
 onto a unique Master curve by a proper scaling of both the $x$ and $y$ axis. We test Eqs. (\ref{univ1},\ref{univ2}) in Fig. \ref{xv} in the case of the Paris Stock 
 Exchange, from which we extract $b \approx 1.02$ and $b' \approx 1.80$. Interestingly, we find that the volume and the
 spread are nearly uncorrelated ($b=1$), whereas the volume traded and the impact are correlated ($b' > 1$), as expected.

 Therefore, using Eq. (\ref{mmmg}) as an equality (as suggested by the empirical results of Section 3), and 
 Eqs. (\ref{dd2},\ref{univ1},\ref{univ2}), we obtain the main result of this section:
 \be
 \label{rgr}
 \langle S\rangle  = c \, \sigma _1,
 \ee
 where $c$ is a stock independent numerical constant, which can be expressed using the constants introduced above 
 as $c=2\lambda b'/\sqrt{a}b$. This very simple relation between volatility {\it per trade} and average 
 spread was noted in \cite{QF04,Zumbach}, and we present further data in the next section to support this conjecture. 
 Therefore, the constraints that (i) optimized high frequency execution strategies impose that the price is diffusive (see 
 \cite{QF04,QF05}), and (ii) 
 the cost of limit and market orders are nearly equal [Eqs.(\ref{cc},\ref{mmmg})], lead to a simple relation between liquidity and volatility. 
 As an important remark, note that the above relation is not expected to hold for the volatility {\it per unit time} 
 $\sigma$, since it involves an extra stock-dependent and time-dependent quantity, namely the the trading frequency $\nu$, 
 through:
 \be\label{sigmanu}
 \sigma =  \sigma_1 \sqrt{\nu}.
 \ee
 We will discuss this issue further in Section 5.

 \subsection{Comparison with empirical data}

\begin{figure}[htbp]
 \begin{center}
 \rotatebox{270}{\resizebox{7.0cm}{!}{\includegraphics{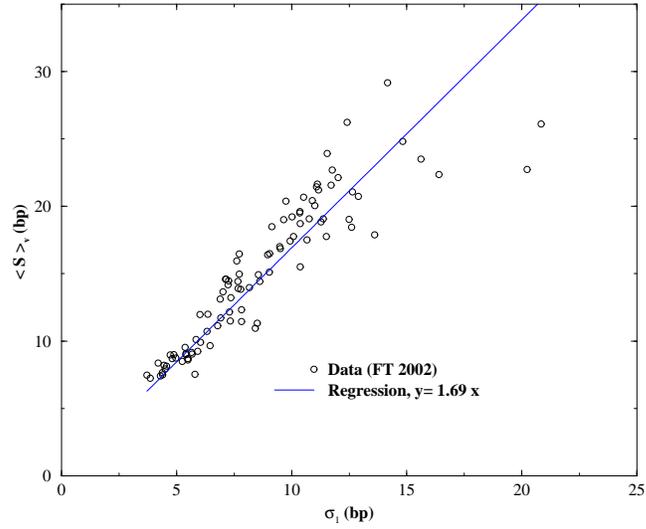}}}
 \caption{\small{Test of Eq. (\ref{rgr}) for France Telecom in 2002. Each point corresponds to a pair 
 ($\langle S \rangle$, $\sigma_1$), computed by averaging over 10000 non overlapping trades ($\sim$ two trading days). 
 Both quantities are expressed in basis points. From a linear fit, we find $c \approx 1.69$ with $R^2=0.90$.}}
 \label{fte}
 \end{center}
 \end{figure}

\begin{figure}[htbp]
 \begin{center}
 \rotatebox{270}{\resizebox{7.0cm}{!}{\includegraphics{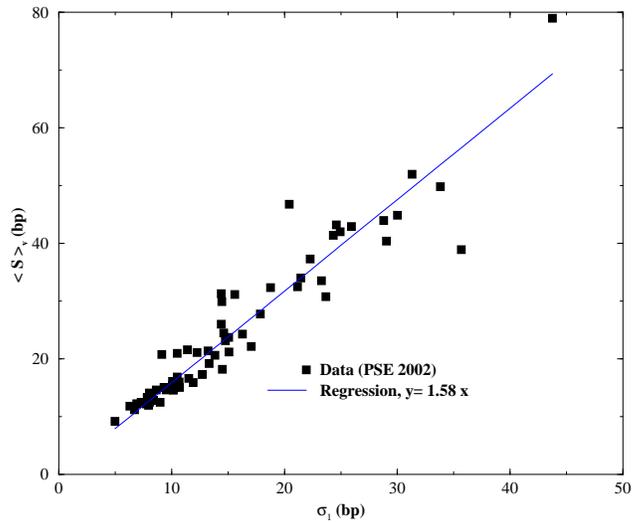}}}
 \caption{\small{Test of Eq. 
 (\ref{rgr}) for 68 stocks from the Paris Stock Exchange in 2002, averaged over the entire year. The value of 
 the linear regression slope is $c \approx 1.58$, with $R^2=0.96$}}
 \label{fte2}
 \end{center}
 \end{figure}

\begin{figure}[htbp]
 \begin{center}
 \rotatebox{270}{\resizebox{7.0cm}{!}{\includegraphics{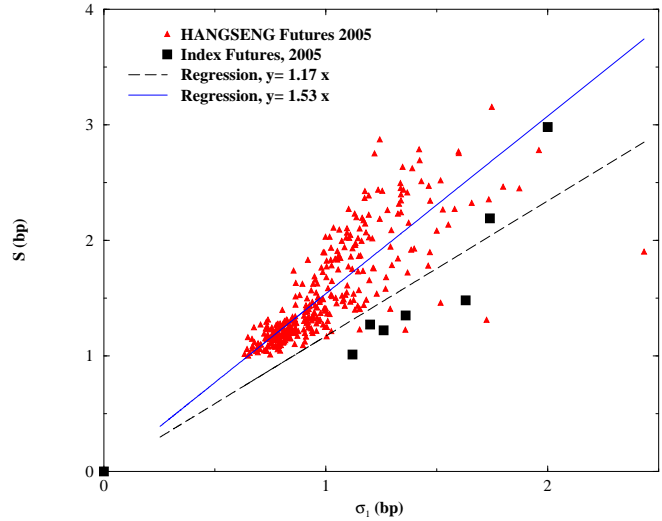}}}
 \caption{\small{Test of Eq. (\ref{rgr}) for the {\sc hangseng} futures contract (triangles), and 
 across small tick Index Futures in 2005: {\sc cac}, {\sc dax}, {\sc ftse}, 
 {\sc ibex}, {\sc mib}, {\sc smi}, {\sc hangseng} (squares). Each point corresponds to a pair 
 ($\langle S \rangle$, $\sigma_1$), computed by averaging either over 1000 non overlapping trades (triangles)
 or over the whole year (squares). From a linear fit, we find $c \approx 1.53$ for the  
 {\sc hangseng} across time and $c \approx 1.17$ across Index Futures.}}
 \label{futs2}
 \end{center}
 \end{figure}

\begin{figure}[htbp]
 \begin{center}
 \rotatebox{270}{\resizebox{7.0cm}{!}{\includegraphics{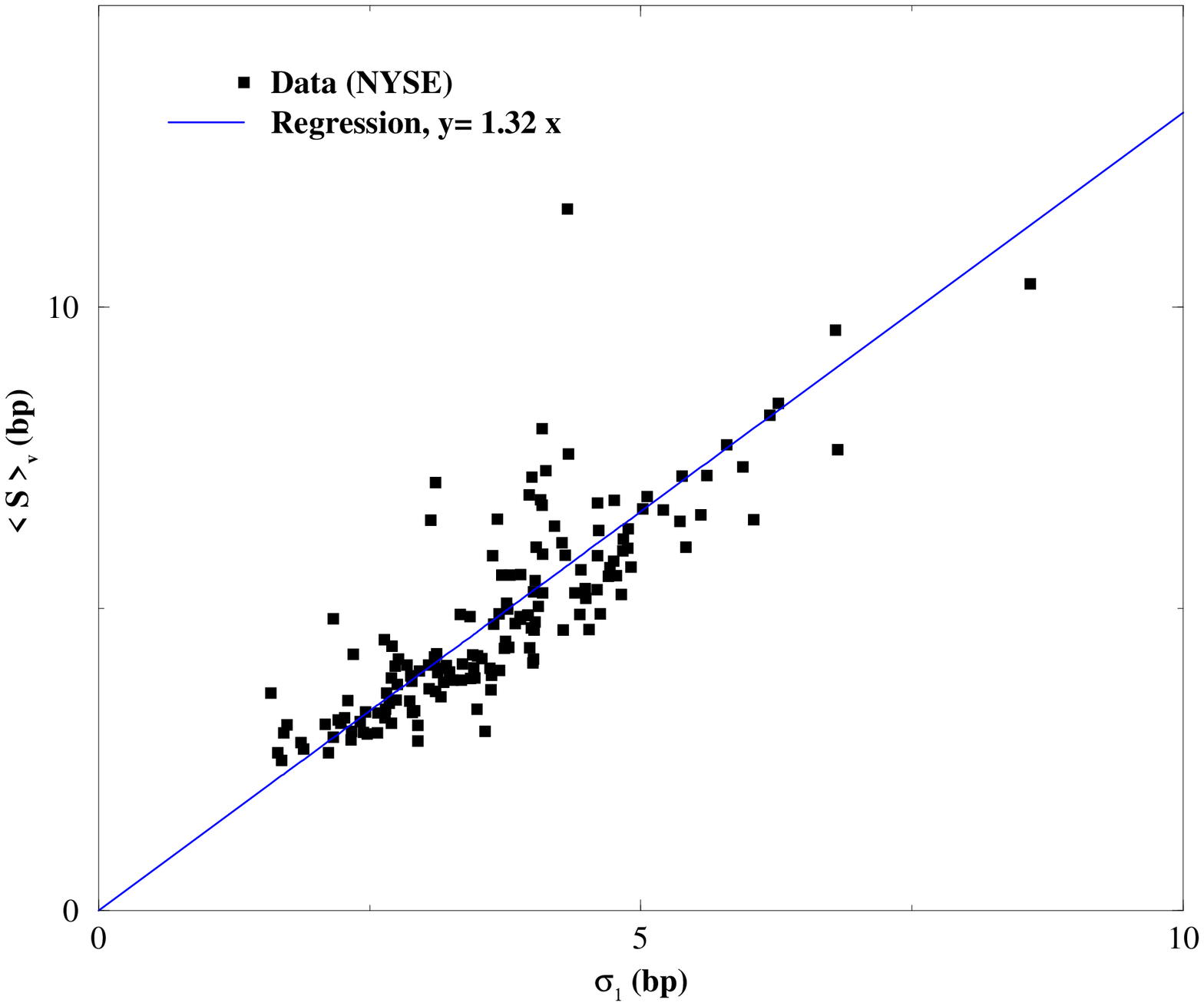}}}
 \caption{\small{Test of Eq. (\ref{rgr}) for  stocks from the {\sc nyse} in 2005. Each point corresponds to a pair 
 ($\langle S \rangle$, $\sigma_1$), computed by averaging over the entire year. 
 Both quantities are expressed in basis points. From a linear fit, we find $c \approx 1.32$, , with $R^2=0.91$}}
 \label{nyse}
 \end{center}
 \end{figure}

 Using the same data sets as in Sections 3.1 and 3.2, we now test empirically the predicted linear relation 
 between spread and volatility per trade, Eq. (\ref{rgr}). The average spread $\langle S \rangle$ is defined as 
 the average distance between bid and ask immediately before each trade (and not as the average over all posted quotes).
 The volatility per trade is defined as the root mean square of the trade by trade return.\footnote{Since prices are very 
 close to random walks, defining the volatility from returns defined on a longer time scale gives very similar results. 
 On our set of {\sc pse} stocks, we find that $\sigma_{128}/\sqrt{128} \approx 0.84 \sigma_1$, indicating a small 
 anti-correlation of returns ($\sim 15 \%$) of short time scales.} Our results for the Paris Stock Exchange are shown 
 in Figs \ref{fte} and \ref{fte2}. We see that Eq. (\ref{rgr}) describes the data very well, with $R^2$s over 0.9. 
 Interestingly, using the
 results obtained above across the {\sc pse} stocks, we have $a' \approx 10.9$, $b \approx 1.02$, $b' \approx 0.53$, 
 $\lambda \approx 1.43$, leading to $c \approx 1.53$, in close correspondence with the direct regression result 
 $c \approx 1.58$. Similar results are obtained for Index futures (Figs. \ref{futs2}-a \& b) or for the {\sc nyse}
 (Fig. \ref{nyse}), with values of $c$ which are all very similar $c \sim 1.2 - 1.6$. We have also checked that there
 is an average intra-day pattern which is followed in close correspondence both by $\langle S \rangle$ and $\sigma_1$:
 spreads are larger at the opening of the market and decline throughout the day. Note that the trading frequency $\nu$ 
 increases as time elapses, which, using Eq. (\ref{sigmanu}), explains the familiar U-shaped pattern of the volatility
 per unit time. 

 \section{Discussion and conclusion}

 The main theoretical result of this paper is the possibility to express the cost of market orders and the 
 profit of infinitesimal market-making/taking strategies in terms of directly observable quantities, namely the spread and the lag-dependent 
 impact function. Imposing that any market taking or liquidity providing strategies is at best marginally profitable 
 allows one to define viable regions of the microstructural ``phase-diagram'' (Fig. 2) where electronic markets should operate, 
 and suggest a linear relation between spread and instantaneous impact. This relation is in good agreement with empirical data on small tick 
 contracts, with a slope compatible with marginal profitability of both fast market making and copy-cat market taking strategies. 
 Somewhat paradoxically, we find that liquidity providers as a whole offer average negative costs to 
 market orders although high frequency 
 market making strategies still manage to get (marginally) compensated. Our analysis allows us to compare in an objective way the spreads 
 in different markets and suggests that spreads are distinctly larger on the {\sc nyse}. Note that our analysis
 does not require any model specific assumptions such as the nature of order flow correlations 
 or the fraction of informed trades. In fact our results hold even if trades were all uninformed but still mechanically impact the 
 price. 

 Making reasonable further assumptions, we have then shown that spread $S$ and volatility per trade $\sigma_1$ 
 are also proportional, a result that we confirm empirically with correlations above $0.9$. This very 
 simple relation means that most of the volatility comes from trading alone, and suggests that the bid-ask 
 spread is dominated by adverse selection, provided one considers the volatility per trade as a measure of 
 the amount of `information' included in prices at each 
 transaction. There are indeed two complementary economic interpretations of the relation $\sigma_1 \sim S$ in small tick markets: 
 \begin{itemize}
 \item (i) since the typical available liquidity in the order book is quite small, market orders tend to grab a significant fraction 
 of the volume at the best price; furthermore, the size of the `gap' above the ask or below the bid is observed to be 
 on the same order of magnitude as the bid-ask spread itself which therefore sets a natural scale for price variations. Hence both the impact and 
 the volatility per trade are expected to be of the order of $S$, as observed; 
 \item (ii) the relation can 
 also be read backward as $S \sim \sigma_1$: when the volatility per trade is large, the risk of placing limit orders is large and therefore
 the spread widens until limit orders become favorable. 
 \end{itemize}
 Therefore, there is a clear two-way feedback that imposes 
 the relation $\sigma_1 \sim S$, valid on average; any significant deviation tends to be corrected by the 
 resulting relative flow of limit and market orders. Our result therefore appears as a fundamental property of the 
 markets organization, which should be satisfied within any theoretical description of the micro-structure. Zero intelligence 
 models \cite{Farmer4}, or bounded-range models \cite{Fou2,Luck,Io} fail to predict any universal relation between $S$ and $\sigma_1$.

 Our relation involves the volatility per trade whereas most of the econometric work has instead 
 focused on the volatility per unit time $\sigma$. The relation between the two involves the trading frequency $\nu$, 
 which is itself both time- and stock-dependent. As a function of time, we find, in agreement with \cite{Engle}, that volatility per trade 
 and trading frequency
 are positively correlated; the volatility $\sigma=\sigma_1 \sqrt{\nu}$ therefore increases because both $\sigma_1$ and
 $\nu$ increase.\footnote{The long-memory property of $\sigma$ is argued in \cite{Gopi} to be related to long range correlation
 in the trading frequency rather than in the volatility per trade, but see \cite{Farmer0}.} Across stocks, on the other hand, the 
 volatility per unit time exhibit only weak systematic variations with capitalization $C$: $\sigma \sim C^\varphi$
 with $\varphi \approx 0$, whereas the trading frequency increases with capitalization as $\nu \sim C^{\zeta}$. 
 For stocks belonging to the {\sc ftse}-100, Zumbach 
 finds $\zeta \approx 0.44$ \cite{Zumbach}, while for US stocks the scaling for $\nu$ is less clear \cite{Eisler}. Interestingly, our result 
 then leads to a result between average spread and capitalization of the form $S \sim C^{\varphi-\zeta/2} \sim 
 {C}^{-0.22}$, in good agreement with Zumbach's data \cite{Zumbach}, with the impact data of Lillo et al. \cite{Lillo} and 
 with our own data on the {\sc pse}.

 The fundamental question at this stage is to know what fixes the volatility $\sigma$ and the trading frequency $\nu$. 
 Clearly, the trading frequency has to do with the available liquidity and the way large volumes have to be cut in small
 pieces. But is the volatility per unit time the {\it primary} object, driven by a fundamental process such as the arrival 
 of news, to which the volatility per trade and therefore the spread is slaved? Or is the market 
 micro-structure and trading activity imposing, in a bottom-up way, the value of the volatility? Understanding these 
 coupled dynamical problems appears to be a major challenge for the theory of financial markets, and an unavoidable step to 
 understand the interrelation between order flow and price changes, and liquidity and market efficiency 
 \cite{MadS,Lyons,Chordia1,Chordia2,Hopman,QF04,Farmer1,Chordia4}.
 \vskip 1cm
 We want to warmly thank S. Bogner, J. D. Farmer, Th. Foucault and G. Zumbach for important and useful discussions. We also thank
 the referees for very constructive remarks, which helped improving the manuscript.

 \section*{Appendix 1: Impact and volatility in the {\sc mrr} model}
 
 From the basic equation determining the dynamics of the mid-point, 
\be
 m_{i+1} - m_i = p_{i+1} - p_i  - \theta \rho (\epsilon_{i}-\epsilon_{i-1})= \xi_i + \theta (1-\rho) \epsilon_i,
\ee
one gets:
\be
m_{i+\ell}-m_i = \sum_{j=i}^{i+\ell-1} \xi_j + \theta (1-\rho) \sum_{j=i}^{i+\ell-1} \epsilon_j.
\ee
Therefore, taking into account the correlation between $\epsilon$s, and the assumption that external 
shocks are uncorrelated with the order flow, the impact function is:
\be
{\cal R}_\ell = \langle \epsilon_i (m_{i+\ell}-m_i) \rangle = \theta (1-\rho) \sum_{\ell'=0}^{\ell-1} \rho^{\ell'}
= \theta (1-\rho^\ell).
\ee
Note that in this model, the `bare' impact function $G_0(\ell)$ defined in \cite{QF04,QF05} through:
\be
m_i = \sum_{j=-\infty}^{i-1} \xi_j +  \sum_{j=-\infty}^{i-1} G_0(i-j-1) \, \epsilon_j,
\ee
is here found to be constant, equal to $G_0(\ell)=\theta (1-\rho)$.
Finally, one finds:
\be
\sigma_1^2 = \langle (m_{i+1} - m_i)^2 \rangle = \Sigma^2 + \theta^2 (1-\rho)^2
\ee
and
\be
\sigma_\infty^2 =  \Sigma^2 + \theta^2 (1-\rho)^2 (1 + 2 \frac{\rho}{1 - \rho}) = \Sigma^2 + \theta^2 (1-\rho^2)
\ee
More generally, assuming that only the sign surprise matters, one can write, for arbitrary 
correlations between signs:
\be
m_{i+\ell}-m_i = \sum_{j=i}^{i+\ell-1} \xi_j + \theta  \sum_{j=i}^{i+\ell-1} \epsilon_j-\langle \epsilon_{j+1} \rangle_j,
\ee
where the last term is the conditional expectation of the next sign. The impact function now generalizes to:
\be
{\cal R}_\ell = \theta \left[ 1 - C(\ell) \right],
\ee
and therefore $\lambda_\infty=1/(1-C_1)$.

 \section*{Appendix 2: Summary statistics}

 \begin{table}
  \begin{center}
  \begin{tabular}{||l|c||c|c||} \hline\hline
  Code\ \hspace{0.05cm} \   & 
  \hspace{0.05cm}  Name \hspace{0.05cm} &
  \hspace{0.05cm} Code \hspace{0.05cm} &
  \hspace{0.05cm} Name \hspace{0.05cm} 
  \\ \hline 
  ACA&Credit Agricole & IFG & Infogrames Entertainment \\  \hline
  AC&Accor &LG&Lafarge\\ \hline
  AF&Air France-KLM& LI&Klepierre\\ \hline 
  AGF&Assurances Generales de France&LY&Suez\\ \hline
  AI&Air Liquide&MC&LVMH \\ \hline
  ALS&Alstom RGPT&ML&	Michelin\\ \hline
  ALT & Altran &MMB&Lagardere\\ \hline
  AVE&Aventis &MMT&M6-Metropole Television\\ \hline
  BB&Societe BIC &NAD&Wanadoo\\ \hline
  BN&Groupe Danone&NK&Imerys\\ \hline 
  CAP&Cap Gemini&OGE&Orange \\ \hline
  CA&Carrefour & OR&L Oreal \\ \hline
  CDI&Christian Dior &PEC&Pechiney\\ \hline
  CGE&Alcatel &PP&PPR \\ \hline
  CK&Casino Guichard (pref.) &PUB&Publicis Groupe\\ \hline
  CL&Credit Lyonnais &RF&Eurazeo\\ \hline
  CNP&CNP Assurances&RHA&Rhodia \\ \hline
  CO&Casino Guichard  &RI&Pernod-Ricard \\ \hline
  CS&AXA &RNO&Renault\\ \hline 
  CU&Club Mediterranee&RXL&Rexel\\ \hline
  CY&Castorama Dubois&SAG&Safran \\ \hline
  DEC&JC Decaux &SAN&Sanofi-Aventis\\ \hline
  DG&Vinci &SAX&Atos Origin\\ \hline
  DSY&Dassault Systemes &SCO&SCOR\\ \hline
  EF&Essilor International&SC&Simco \\ \hline
  EN&Bouygues& SU&Schneider Electric\\ \hline
  FP&Total &SW&Sodexho Alliance\\ \hline
  FR&Valeo &TEC&Technip\\ \hline
  FTE&France Telecom&TFI&Television Francaise 1\\ \hline
  GFC&Gecina & TMM&Thomson\\ \hline
  GLE&Societe Generale&UG&Peugeot \\ \hline
  GL&Galeries Lafayette &UL&Unibail\\ \hline
  HAV&Havas &VIE&Veolia Environnement\\ \hline
  HO&Thales & ZC & Zodiac \\ \hline
  \end{tabular}
  \end{center}
  \caption[]{\small Codes and names of the {\sc pse} stocks analyzed in Table 2.}
  \end{table}

 \begin{table}
 \begin{center}
 \begin{tabular}{||l|c|c|c|c|c|c|c|c|c|c|c||} \hline\hline
 Code\ \hspace{0.05cm} \   & 
 \hspace{0.05cm}  Tnv \hspace{0.05cm} &
 \hspace{0.05cm} $\langle q_t \rangle$ \hspace{0.05cm} &
 \hspace{0.05cm} $\langle v \rangle$ \hspace{0.05cm} &
 \hspace{0.05cm} \#  \hspace{0.05cm} &
 \hspace{0.05cm} $\sigma_1$ \hspace{0.05cm} &
 \hspace{0.05cm} $\langle S \rangle$ \hspace{0.05cm} &
 \hspace{0.05cm} $\sigma_S$ \hspace{0.05cm} &
 \hspace{0.05cm} ${\overline{\cal R}}_1$ \hspace{0.05cm} &
 \hspace{0.05cm} $C_1$ \hspace{0.05cm} &
 \hspace{0.05cm} $\lambda_\infty$ \hspace{0.05cm} &
 \hspace{0.05cm} Tick \hspace{0.05cm} 
 \\ \hline
 ACA & 15.4 & 78 & 11.1 & 347 & 9.12 & 20.76 & 18.28 & 2.30 & 0.28 & 1.70 & 5.13 \\ \hline
 AC & 22.1 & 87 & 26.3 & 211 & 8.97 & 12.45&  13.00&  2.99&  0.26 & 2.16 & 2.68 \\ \hline
 AF & 4.0 &40& 7.2 &139& 14.40 &25.96& 25.57 &3.85 & 0.23 & 2.30 &  7.01 \\ \hline
 AGF & 9.6 &80 &16.9 &142& 13.32 &19.17 &18.88& 4.28& 0.28  & 2.28 & 5.46 \\ \hline
 AI & 34.0 &154 &26.9 &316& 8.21& 13.72& 11.07 &3.06&  0.26 &  1.42 & 6.87 \\ \hline
 ALS & 14.2& 57 &13.0 &274 &14.75& 23.11 &20.80& 4.84 & 0.14  & 1.84 & 12.02 \\ \hline
 ALT & 12.7& 53& 14.6 &217 &23.66& 30.75 &27.88 &7.15& 0.17 & 2.71 & 11.04 \\ \hline
 AVE & 134.1 &362 &62.6& 536& 6.70 &11.18& 7.31 &2.59&  0.21 & 1.72 & 7.54 \\ \hline
 BB  &0.8 &30& 11.9 &16 &28.80& 43.94& 47.51 &7.78 & 0.33 & 1.66 & 2.55 \\ \hline
 BN & 59.5 &310& 45.9& 324& 6.30 &11.80& 7.79& 2.37 & 0.28 & 1.03 & 7.62 \\ \hline
 CAP & 34.7 &94 &28.9 &300 &12.71& 17.27& 17.03 &4.39& 0.20 & 2.55 &5.17 \\ \hline
 CA & 79.8 &229 &40.1 &497 &6.92 &12.20 &9.16 &2.48 & 0.22 & 1.85 & 5.35 \\ \hline
 CDI & 5.1 &59 &14.9 &86 &18.76 &32.32 &29.08 &6.23 & 0.29 & 1.74 & 2.79 \\ \hline
 CGE & 79.8 &152 &21.3 &936 &10.49 &20.94 &14.85 &3.31 & 0.13 &  1.70 & 15.37 \\ \hline
 CK & 0.3 &37 &11.4 &6 & 20.43 &46.72 &32.24& 7.37 & 0.22 & 0.93 & 8.06 \\ \hline
 CL & 30.1 &163 &23.0 &328& 8.05& 13.60& 13.04 &1.96 & 0.36 & 1.77 & 3.43 \\ \hline
 CNP&  1.7& 41 &9.3& 45& 15.58& 31.12& 33.74 &3.53& 0.31 & 1.43 & 2.67 \\ \hline
 CO & 13.8& 98 &22.9 &151& 10.38& 15.43& 14.42& 3.47&  0.27 & 1.50 & 6.58 \\ \hline
 CS & 96.1 &144 &36.6& 657 &8.32& 12.78& 10.17& 3.13& 0.17 & 1.97 & 6.39 \\ \hline
 CU & 1.0 &22 &7.7& 32& 30.02& 44.83& 47.65& 8.85 & 0.23 & 1.84 & 3.97 \\ \hline
 CY & 15.4& 2148& 77.1 &50 &8.64& 14.61& 14.49& 2.35 & 0.26 & 1.67 & 8.03 \\ \hline
 DEC&  0.7 &27 &12.7& 13& 43.75& 78.92& 77.84& 13.81&  0.33 & 1.53 & 8.15 \\ \hline
 DG & 16.8 &134& 25.4& 165 &8.02& 14.05 &10.90 &2.76 & 0.22 & 1.24 & 7.76 \\ \hline
 DSY&  10.4 &65 &19.6& 133 &17.06& 22.11& 23.83& 5.71& 0.27 & 2.30 & 5.28 \\ \hline
 EF & 5.6& 59 &20.4& 69& 13.21& 21.38& 23.01& 3.94 &0.25 & 1.90 & 2.52 \\ \hline
 EN & 17.8 &66 &21.3& 210& 10.15 &14.55 &14.35& 3.47&  0.25 & 1.85 & 3.46 \\ \hline
 FP & 322.4& 662& 114.4& 705& 4.96& 9.17& 4.86& 2.12 &0.20 & 1.49 & 6.71 \\ \hline
 FR & 8.3 &80& 24.1& 86& 13.84& 20.60& 22.46& 4.39& 0.25 & 1.99 & 3.50 \\ \hline
 FTE& 112.7& 137 &29.5& 956& 9.12& 15.04& 12.57& 2.93&  0.14 & 1.85 & 6.39 \\ \hline
 GFC&  0.3 &35& 9.9& 7 &17.85& 27.74& 31.55& 4.56& 0.27 &  0.98 & 5.62 \\ \hline
 GLE&  82.7& 239& 46.0& 449& 8.07& 12.70& 9.27 &3.04& 0.23 & 2.26 & 7.07 \\ \hline
 GL & 0.9 &43 &13.6 &17 &24.34 &41.34 &39.96 &6.92 & 0.29 & 1.38 & 7.24 \\ \hline
 HAV&  8.1 &57 &14.6 &139 &21.15 &32.44 &26.53 &7.41 & 0.18 & 1.91 & 17.15 \\ \hline
 HO & 11.3 &61& 19.7 &143 &10.68 &15.02 &15.86 &3.65 & 0.29 & 1.94 & 2.85 \\ \hline 
 IFG&  3.2 &24 &4.9 &163 &29.05 &40.35 &32.95 &8.51 &0.15 & 2.31 & 21.79 \\ \hline
 LG & 38.2 &193 &36.6& 261 &7.82 &13.31 &9.99 &2.90 &0.24 & 1.67 & 7.33 \\ \hline
 LI & 0.2 &50 &15.8 &3& 14.43& 29.90& 23.18& 4.69 &0.20 & 0.70 & 8.54 \\ \hline
 \hline
 \end{tabular}
 \end{center}
 \end{table}

 \begin{table}
 \begin{center}
 \begin{tabular}{||l|c|c|c|c|c|c|c|c|c|c|c||} \hline\hline
 Code\ \hspace{0.05cm} \   & 
 \hspace{0.05cm}  Tnv \hspace{0.05cm} &
 \hspace{0.05cm} $\langle q_t \rangle$ \hspace{0.05cm} &
 \hspace{0.05cm} $\langle v \rangle$ \hspace{0.05cm} &
 \hspace{0.05cm} \#  \hspace{0.05cm} &
 \hspace{0.05cm} $\sigma_1$ \hspace{0.05cm} &
 \hspace{0.05cm} $\langle S \rangle$ \hspace{0.05cm} &
 \hspace{0.05cm} $\sigma_S$ \hspace{0.05cm} &
 \hspace{0.05cm} ${\overline{\cal R}}_1$ \hspace{0.05cm} &
 \hspace{0.05cm} $C_1$ \hspace{0.05cm} &
 \hspace{0.05cm} $\lambda_\infty$ \hspace{0.05cm} &
 \hspace{0.05cm} Tick \hspace{0.05cm} 
 \\ \hline
 LY & 58.4 &111 &28.8 &507& 8.40 &12.84& 11.94 &3.11&  0.20 & 1.97 & 4.31 \\ \hline
 MC & 52.9& 143& 34.7 &381& 8.01& 12.41& 10.75 &3.01 &0.26 & 2.20 & 4.18 \\ \hline
 ML & 13.2 &71 &21.2 &156& 10.23 &14.98 &15.36 &3.35 &0.30 & 1.93 & 2.71 \\ \hline
 MMB&  13.3 &76 &18.9& 177& 10.67 &15.96 &15.84 &3.60 & 0.25 & 1.94 & 3.50 \\ \hline
 MMT&  0.7 &33 &10.1 &17 &33.82 &49.77 &48.42 &9.71 &0.32 & 1.66 & 3.57 \\ \hline
 NAD&  4.6 &47 &6.0 &188 &14.39 &31.27 &20.91 &4.11 & 0.17 & 1.89 & 20.36 \\ \hline
 NK & 1.1 &43 &13.4 &21 &25.95 &42.89 &39.34 &7.69 &0.29 & 1.40 & 8.07 \\ \hline
 OGE&  36.4 &182 &21.2 &429 &11.40& 21.53 &13.35 &3.57 &0.16 & 1.82 & 16.03 \\ \hline
 OR & 68.0 &211 &41.9 &406& 7.30& 12.45 &9.45 &2.76 &0.25 &  1.39 & 6.59 \\ \hline
 PEC&  9.5 &110 &31.9 &75 &15.06 &23.71& 24.89 &5.07 & 0.24 & 2.41 & 5.45 \\ \hline
 PP & 36.8 &154 &31.5 &292 &10.08 &15.44 &13.25& 3.39&  0.25 & 2.23 & 7.23 \\ \hline
 PUB&  11.9 &78 &27.3 &109 &15.08 &21.19 &22.92 &4.97 &0.23 & 2.46 & 3.85 \\ \hline
 RF & 0.1 &25 &7.0 &3 &22.27& 37.26 &32.69 &7.07 &0.23 & 0.72 & 8.25 \\ \hline
 RHA&  2.1 &32 &9.7 &55 &21.45 &33.99 &32.97 &6.71 & 0.22 & 1.87 & 10.83 \\ \hline
 RI & 12.6 &138 &39.1 &80& 10.49& 16.82 &16.01& 3.64& 0.22 & 1.49 & 6.42 \\ \hline
 RNO&  35.8 &158 &34.4 &260 &8.01 &12.80 &11.13& 2.71 & 0.25 & 2.27 & 4.38 \\ \hline
 RXL&  0.7 &34 &13.0 &14& 31.30 &51.91 &50.49 &9.21 &0.26 & 1.50 &  6.77 \\ \hline
 SAG&  1.5 &36 &10.3 &35 &24.59 &43.15 &42.89 &7.29 & 0.22 & 1.73 & 7.48 \\ \hline
 SAN&  94.2 &301 &56.4 &417& 7.76& 12.18& 8.60 &3.04 &0.25 & 1.49 & 7.92 \\ \hline
 SAX&  6.0 &57& 20.4 &73& 23.28& 33.48& 33.79 &7.45 & 0.27 & 2.58 & 5.23 \\ \hline
 SCO& 2.0 &25 &9.3 &55 &35.67& 38.88 &40.00 &8.37 &0.23 & 2.16 &7.40 \\ \hline
 SC &0.5 &55& 13.9 &10 &12.24 &21.08 &19.01 &3.58 &0.22 & 1.28 & 6.10 \\ \hline
 SU &26.2 &129 &33.0 &198 &9.52 &14.60 &13.43 &3.30 &0.26 & 2.15 & 5.63 \\ \hline
 SW &11.2 &67 &19.5& 144 &14.48 &18.19& 20.40 &4.26 &0.32 & 1.92 &  3.22 \\ \hline
 TEC& 9.4 &123 &31.6 &74 &16.27 &24.27 &25.77 &5.06 &0.24 &  2.18 &  7.31 \\ \hline
 TFI& 17.4 &63 &21.0 &207 &11.91 &15.87&16.19 &4.09 &0.25 & 2.07 & 3.71 \\ \hline
 TMM& 28.0 &78 &20.8 &338 &10.08 &16.08 &15.59 &3.18 & 0.19  &2.34 & 4.29 \\ \hline
 UG &33.2 &141 &36.2 &229 &7.95 &11.91 &10.80 &2.86 &0.26 & 2.11 & 4.43 \\ \hline
 UL &3.0 &64 &22.9 &33 &14.61 &24.45 &23.47 &4.87&0.27 & 1.41 & 8.07 \\ \hline
 VIE& 19.8 &77 &24.8 &199 &11.52 &16.60 &17.81 &3.75 &0.23 & 2.23 & 3.57 \\ \hline
 ZC &0.9 &33 &8.7 &26 &24.95 &41.99 &42.40 &7.28&0.24 & 1.64 &4.21 \\ \hline\hline
 \end{tabular}
 \end{center}
 \caption[]{\small Pool of the 68 stocks of the {\sc pse} studied in this paper, with their
 summary statistics for 2002. The daily turnover is in million Euros, $\langle q_t \rangle$ is the
 average amount in book (bid+ask) in thousand Euros, $\langle v \rangle$ is the 
 average size of market order (in thousand Euros). The total number of trades (in thousand) corresponds to the whole 
 year 2002. The volatility per trade $\sigma_1$, the average spread $\langle S \rangle$, the spread standard deviation
 $\sigma_S$, the average response ${\overline{\cal R}}_1$ and the average tick size are all in basis points. Note that $\sigma_S 
 \approx \langle S \rangle$, characteristic of an exponential distribution of the spread. Note also that the volume 
 available at the best prices is $\sim 10^{-3}$ of the daily turnover. 
 }
 \end{table}

 \begin{table}
 \begin{center}
 \begin{tabular}{||l|c|c|c|c|c|c|c|c|c||} \hline\hline
 Code\ \hspace{0.05cm} \   & 
 \hspace{0.05cm}  Turnover \hspace{0.05cm} &
 \hspace{0.05cm} $\langle q_t \rangle$ \hspace{0.05cm} &
 \hspace{0.05cm} $\langle v \rangle$ \hspace{0.05cm} &
 \hspace{0.05cm} \# trade \hspace{0.05cm} &
 \hspace{0.05cm} $\sigma_1$ \hspace{0.05cm} &
 \hspace{0.05cm} $\langle S \rangle$ \hspace{0.05cm} &
 \hspace{0.05cm} $\sigma_S$ \hspace{0.05cm} &
 \hspace{0.05cm} ${\overline{\cal R}}_1$ \hspace{0.05cm} &
 \hspace{0.05cm} Tick \hspace{0.05cm} 
 \\ \hline
 LMEB & 262 & 21 & 199 &211 &8.4 &46.7 &4.4 &11.5 &46.6 \\ \hline \hline
 \end{tabular}
 \end{center}
 \caption[]{\small Summary statistics for Ericsson in the period March 2004-November 2004. 
 Units are the same as in Table 2, except $\langle q_t \rangle$ which is now in million Euros.
 Note that $\langle v \rangle/\langle q_t \rangle \approx 10^{-2}$.}
 \end{table}


\newpage

 \begin{thebibliography}{99}

 \bibitem{Orlean} A. Orl\'ean, {\it Le pouvoir de la finance}, Odile Jacob, Paris (1999);
 {\it A quoi servent les march\'es financiers ?}, in {Qu'est-ce que la Culture ?}, 
 Odile Jacob, Paris (2001).

 \bibitem{OH} M. O'Hara, {\it Market microstructure theory}, Cambridge, Blackwell, (1995).

 \bibitem{BFH} B. Biais, Th. Foucault, P. Hillion, {\it Microstructure des march\'es financiers}, PUF (1997)

 \bibitem{Mad} A. Madhavan, {\it Market microstructure: A survey}, Journal Financial Markets, {\bf 3}, 205 (2000)

 \bibitem{Glo1} L. R. Glosten, P. Milgrom, {\it Bid, Ask and Transaction Prices in a Specialist Market with 
 Heterogeneously Informed Traders}, Journal of Financial Economics {\bf 14} 71 (1985) 

 \bibitem{Glo2} L. R. Glosten, {\it Components of the Bid-Ask Spread and the Statistical Properties of Transaction Prices}, 
 Journal of Finance, {\bf 42} 293 (1987).

 \bibitem{Has} J. Hasbrouck, {\it Measuring the information content of stock trades}, Journal of Finance, {\bf XLVI}, 179 (1991). 

 \bibitem{HS} R. D. Huang, H. R. Stoll, {\it The components of the Bid-Ask spread: A general approach}, Review of Financial
 Studies, {\bf 4} 995 (1997).

 \bibitem{MadS} A. Madhavan, M. Richardson, M. Roomans, {\it Why do security prices fluctuate? a transaction-level analysis of 
 NYSE stocks}, Review of Financial Studies, {\bf 10}, 1035 (1997).
 
 \bibitem{Holli} B. Hollifield, A. Miller, P. Sandas, {\it Empirical Analysis of Limit Order Markets}, 
 forthcoming Review of Economic Studies.
 
 \bibitem{Stoll} H. Stoll, {\it Friction}, Journal of Finance , {\bf 55} 1479 (2000)

 \bibitem{Black} F. Black, {\it Towards a fully automated exchange}, Review Financial Analysts, {\bf 27}, 29 (1971).

 \bibitem{Kyle} A. S. Kyle, {\it Continuous auctions and insider trading},  Econometrica {\bf 53}, 1315 (1985).

 \bibitem{Liu} W.-M. Liu, {\it Monitoring and Limit Order Submission Risks}, 
 University of New South Wales Working Paper, 2005.

 \bibitem{QF04} J. P. Bouchaud, Y. Gefen, M. Potters, M. Wyart, {\it  Fluctuations and response in financial markets: The
 subtle nature of `random' price changes},  Quantitative Finance {\bf 4}, 176 (2004).

 \bibitem{Farmer1} F. Lillo, J. D. Farmer, {\it The long memory of efficient markets}, Studies in Nonlinear Dynamics and Econometrics, {\bf 8}, 1 (2004).

 \bibitem{QF05} J. P. Bouchaud, J. Kockelkoren, M. Potters, {\it Random walks, liquidity molasses and critical response 
 in financial markets}, Quantitative Finance, {\bf 6}, 115 (2006).

 \bibitem{Bess} H. Bessembinder, {\it Bid-Ask Spread in the Interbank Foreign Exchange Markets}, Journal of Financial Economics, {\bf 35},
 317 (1994)

\bibitem{Cop} M. Coppejans, I. Domowitz, A. Madhavan, {\it Liquidity in Automated Auction}, Working paper, 2001.

\bibitem{Chordia1} T. Chordia, R. Roll, A. Subrahmanyam, {\it Order Imbalance, Liquidity and Market Returns}, Journal of Financial
 Economics, {\bf 65}, 111 (2001).

\bibitem{Chordia2} T. Chordia, A. Subrahmanyam, {\it Order imbalance and individual stock returns}, Journal of Financial Economics, {\bf 72} 
 485 (2004)

\bibitem{Chordia3} T. Chordia, L. Shivakumar, A. Subrahmanyam, {\it Liquidity dynamics across small and large firms}, 
 Economic Notes, {\bf 33} 111  (2004)
 
\bibitem{Farmer0} L. Gillemot, J. D. Farmer, F. Lillo, {\it There's more to volatility than volume}, physics/0510007 

\bibitem{Zumbach} G. Zumbach, {\it How the trading activity scales with the company sizes in the {\sc ftse} 100}, Quantitative Finance,
 {\bf 4}, 441 (2004)

\bibitem{Fou1} Th. Foucault, {\it Order flow composition and trading costs in a dynamic limit order market}, Journal of Financial Market, 2, 99 (1999)

\bibitem{Fou2} Th. Foucault, O. Kadan, E. Kandel, {\it Limit order book as a market for liquidity}, Review
of Financial Studies, 18, 1171 (2003).

\bibitem{Farmer2} M. G. Daniels, J. D. Farmer, G. Iori, E. Smith,
 {\it Quantitative model of price diffusion and market friction based on 
 trading as a mechanistic random process}, Phys. Rev. Lett. {\bf 90}, 108102 (2003).  

\bibitem{Farmer3} E. Smith, J. D. Farmer, L. Gillemot, S. Krishnamurthy, {\it
 Statistical theory of the continuous double auction}, Quantitative Finance {\bf 3}, 481 (2003).

\bibitem{BMP} J.P. Bouchaud, M. M\'ezard, M. Potters, {\it Statistical properties of stock order books: empirical results 
 and models}, Quantitative Finance {\bf 2}, 251 (2002).

\bibitem{Luck} H. Luckock, {\it A steady-state model of the continuous double auction}, Quantitative Finance, {\bf 3}, 385 (2003)

\bibitem{Io} I. Rosu, {\it A dynamic model of the limit order book}, Working paper, 2005.

\bibitem{Farmer4} J. D. Farmer, P. Patelli and I. Zokvo, {\it The predictive power of zero intelligence in financial 
 markets}, PNAS {\bf 102}, 2254-2259 (2005) 

\bibitem{Mike} S. Mike, J. Doyne Farmer, {\it An empirical behavioral model of price formation}, physics/0509194.

\bibitem{Weber} P. Weber, B. Rosenow, {\it Order book approach to price impact}, Quantitative Finance, {\bf 5}, 357 (2005)

\bibitem{FarmerL} F. Lillo, S. Mike, and J. D. Farmer, {\it Theory for long memory in supply and demand}, 
Physical Review E, 7106, 287 (2005).

\bibitem{Almgren} R. Almgren, C. Thum, E. Hauptmann, H. Li, {\it Direct estimation of equity market impact}, 
 working paper (May 2005).
 
\bibitem{Hopman} C. Hopman, {\it Are supply and demand driving stock prices?}, MIT working paper, Dec. 2002, to 
 appear in Quantitative Finance.
 
\bibitem{Handa} P. Handa, R. A. Schwartz, A. Tiwari, {\it The ecology of an order-driven market}, Journal of Portfolio
Management, Winter (1998), 47-56.

\bibitem{HH} L. Harris, J. Hasbrouck, {\it Market vs. limit orders: the SuperDOT
evidence on order submission strategy}, Journal of Financial and Quantitative Analysis 31, 213-31 (1996).

\bibitem{HandaS} P. Handa, R. A. Schwartz, {\it Limit order trading}, J. Finance, {\bf 51}, 1835 (1996).

\bibitem{Bessembinder02} see e.g. H. Bessembinder, {\it Issues in assessing trade execution costs}, 
Journal of Financial Markets {\bf 6}, 233 (2003), and refs. therein. 

\bibitem{Farmer5} J. D. Farmer, L. Gillemot, F. Lillo, S. Mike, A. Sen, 
 {\it What really causes large price changes?}, Quantitative Finance, {\bf 4}, 383 (2004).

\bibitem{Lyons} M. D. Evans, R. K. Lyons, {\it Order Flow and Exchange Rate Dynamics}, Journal of Political Economy, 
{\bf 110} 170 (2002) 


\bibitem{Rosenow} B. Rosenow, {\it Fluctuations and market friction in financial trading}, Int. J. Mod. Phys. C, 
 {\bf 13}, 419 (2002).
    
\bibitem{Lillo} {F. Lillo, J. D. Farmer, R. Mantegna}, {\it Master curve for price-impact function}, Nature, {\bf 421}, 129 (2003). 

\bibitem{Gopi2} {V. Plerou, P. Gopikrishnan, X. Gabaix, H. E. Stanley}, 
{\it Quantifying Stock Price Response to Demand Fluctuations}, Phys. Rev. E 
{\bf 66}, 027104 (2002). 
     	
\bibitem{Gopi} V. Plerou, P. Gopikrishnan, L. A. N. Amaral, X. Gabaix, and H. E. Stanley, {\it Diffusion and Economic Fluctuations}, Phys. Rev. E {\bf 62}, 
3023 (2000). 

\bibitem{Engle} R. F. Engle, {\it The econometrics of ultra-high frequency data}, Discussion paper 96-15,
University of California, San Diego, 1996.

\bibitem{Eisler} Z. Eisler, J. Kertecz, {\it Size matters, some stylized facts of the market revisited}, xxx.lanl.gov/
physics/0508156.

\bibitem{Chordia4} T. Chordia, R. Roll, A. Subrahmanyam, {\it Liquidity and Market Efficiency}, Working paper 2005.

\end{thebibliography}
\end{document}